%% file: 0-main.tex
\documentclass[article]{IEEEtran}
\usepackage{cite}
\usepackage{amsmath,amssymb,amsfonts}
\usepackage{graphicx}
\usepackage{textcomp}

\usepackage{hyperref}
\usepackage{times}  
\usepackage{helvet}  
\usepackage{courier}  
\usepackage{url}  
\usepackage{graphicx}  
\usepackage{subfig}
\usepackage{multirow}
\usepackage{booktabs}
\usepackage{mathrsfs}
\usepackage{algorithm}
\usepackage{color}
\usepackage{comment}
\usepackage{amssymb}
\usepackage{amsmath}
\usepackage{booktabs}
\usepackage[table,xcdraw]{xcolor}
\usepackage{algpseudocode}

\newcommand{\ff}[1]{{\color{black}{#1}}}
\newcommand{\xc}[1]{{\color{black}{#1}}}
\newcommand{\xcc}[1]{{\color{black}{#1}}}

\def\ie{\emph{i.e.}}
\def\eg{\emph{e.g.}}

\def\etc{{\em etc.}} 
\newcommand{\para}[1]{\vspace{.05in}\noindent\textbf{#1}}
\definecolor{newcolor}{rgb}{.8,.349,.1}

\begin{document}
\title{Robust Medical Image Classification from Noisy Labeled Data with Global and Local Representation Guided Co-training}
\author{Cheng Xue, Lequan Yu, Pengfei Chen, Qi Dou, and Pheng-Ann Heng 
\thanks{C. Xue, P. Chen, Q. Dou and P.A. Heng are with the Department of Computer Science and Engineering, The Chinese University of Hong Kong,
Hong Kong, China (e-mail: xchengjlu@gmail.com).}
\thanks{L. Yu is with the Department of Statistics and Actuarial Science, The University of Hong Kong, Hong Kong, China (email:lqyu@hku.hk). (Corresponding author: Lequan Yu.)}
}
\maketitle

\begin{abstract}
Deep neural networks have achieved remarkable success in a wide variety of natural image and medical image computing tasks. However, these achievements indispensably rely on accurately annotated training data. 
If encountering some noisy-labeled images, the network training procedure would suffer from difficulties, leading to a sub-optimal classifier. 
This problem is even more severe in the medical image analysis field, as the annotation quality of medical images heavily relies on the expertise and experience of annotators. 
In this paper, we propose a novel collaborative training paradigm with global and local representation learning for robust medical image classification from noisy-labeled data to combat the lack of high quality annotated medical data.
Specifically, we employ the self-ensemble model with a noisy label filter to efficiently select the clean and noisy samples.
Then, the clean samples are trained by a collaborative training strategy to eliminate the disturbance from imperfect labeled samples.
Notably, we further design a novel global and local representation learning scheme to implicitly regularize the networks to utilize noisy samples in a self-supervised manner. 
We evaluated our proposed robust learning strategy on four public medical image classification datasets with three types of label noise, \ie, random noise, computer-generated label noise, and inter-observer variability noise.
%
%
Our method outperforms other learning from noisy label methods and we also conducted extensive experiments to analyze each component of our method.
\end{abstract}


\input{1-introduction}
\input{2-relatedwork}

\input{3-method}

\input{4-experiment}
\input{5-discussion}



\bibliographystyle{IEEEtran}
\bibliography{ref}

\end{document}

%% file: 1-introduction.tex
\section{Introduction}
\begin{figure}[t]
\centering
  \includegraphics[width=0.48\textwidth]{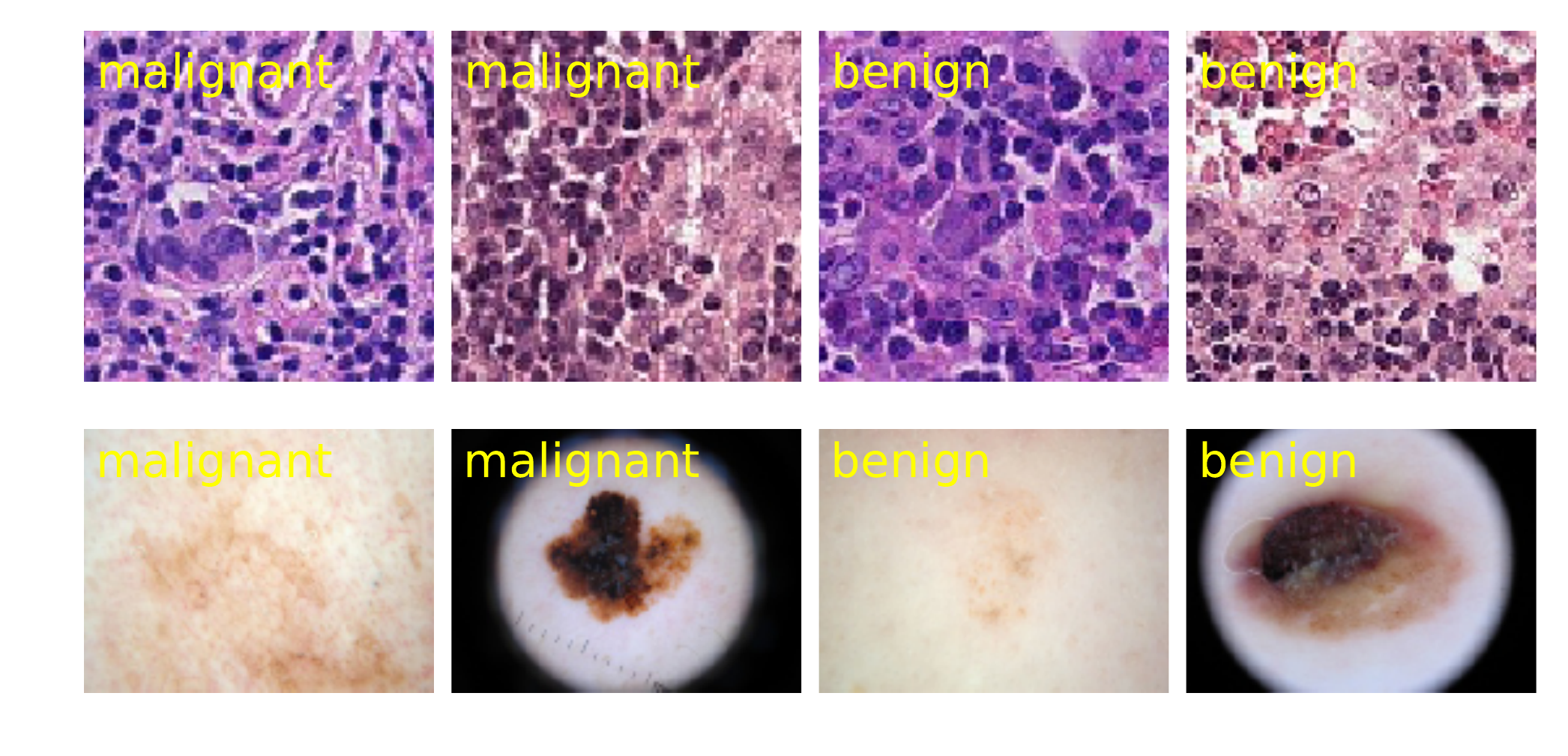} 
\caption{Typical examples of ambiguous malignant and benign images from histology lymph node data ({\textbf{Top}}) and skin lesion data ({\textbf{Bottom}}), which are easily to be incorrectly labeled in the practical annotation process.}
\label{fig:intro}
\vspace{-8pt}
\end{figure}
Robust medical image analysis has been an increasingly important topic, as accurate and robust algorithms are desired to increase clinical workflow efficiency and reliably support medical-decision-makings.
With the advent of deep learning, encouraging human-level performance has been achieved in a spectrum of challenging medical image diagnosis applications including skin cancer~\cite{esteva2017dermatologist}, lung cancer~\cite{setio2017validation}, retinal diseases~\cite{de2018clinically}, histology image analysis~\cite{bejnordi2017diagnostic}, \etc. The success of these applications relies on highly discriminative representations learned by convolutional neural networks (CNNs) from a large amount of carefully labeled data with the aid of domain experts.

Despite the remarkable success, it has been frequently seen that the performance of CNNs is easily affected due to the bias of the training set in complex real-world situations. For instance, such bias could come from statistical distribution mismatch across different clinical sites, imbalance of the number of samples across different disease categories, and the variations of disease visual patterns across patient age or gender populations.
Notably, besides these data biases in the aspect of the image itself, the bias with annotation quality commonly exists, \ff{and is} harmful to the supervised-learning-based diagnostic solutions.
One typical issue is the problem of noisy labels associated with the collected data, especially for the ambiguous medical images which may confuse clinical experts. 
For example, as illustrated in Fig.~\ref{fig:intro}, the malignant histology lymph node images present quite similar colors and structures to benign ones, which shows that noisy annotations are inevitable in medical image analysis in practice. 
The same situation can also be observed for skin lesion images.
Moreover, automatically extracting labels from radiological reports by Natural Language Processing (NLP) tools will also result in label noise. 
Therefore, tackling the annotation \ff{noise} is an important topic in the medical image analysis field.
The straightforward solution is manually reducing the presence of incorrect labels. However, this procedure is expensive, time-consuming and impractical.
The naive training paradigm with noisy samples is reported to be inadequate, since the CNNs are parameterized with a large model capacity and can eventually memorize all training sample pairs including wrongly-labeled ones~\cite{zhang2016understanding,chen2019understanding}. 

Although handling noisy labeled medical data is a crucial task in the era of deep learning, merely preliminary studies have been conducted in the medical image analysis field. 
As far as we know, Dagni et al. \cite{dgani2018training} developed a robust training strategy by explicitly modeling the label noise as a part of network architecture, and presented the application of binary classification for breast mammograms.
Xue et al.~\cite{xue2019robust} proposed to robustly train a CNN classifier with online sample mining and re-weighting approach based on the model uncertainty. Zhu et al. \cite{zhu2019pick} proposed \ff{an} automatic quality evaluation module and overfitting control module to update the network parameters. Shu et al.\cite{shu2019lvc} presented \ff{a novel loss} function by combining noisy labels with image local visual cues to generate better semantic segmentation. Xue et al. \cite{xue2020cascaded} proposed a joint optimization scheme with label correction \ff{module}. 
In more broad literature on deep learning, the topic of noisy label learning has been actively investigated in natural image processing. 
Some methods focus on the estimation of the latent transition matrix of labels and use it to calibrate the loss of noisy samples.
For example, Goldberger et al.~\cite{goldberger2016training} estimated the transition matrix of noisy labels by adding an extra softmax operation on top of the original softmax layer. Patrini et al.~ \cite{patrini2017making} used a two-step solution that firstly computes the label transition matrix then \ff{corrected} the errors for suspicious noisy samples. These methods rely heavily on estimating the noisy class posterior. However, the estimation error for noisy class posterior could be largely due to the randomness of label noise and the complicated data biases, which would lead the transition matrix to be poorly estimated. 

To be free of estimating noisy label transition matrix, a promising direction is to train deep learning models with clean samples, which can be selected according to the noisy classifier's prediction~\cite{zhengerror}. 
Intuitively, if the noisy samples could be automatically filtered out from the training database during learning, the model robustness on noisy labeled data is gained. For instance, Jiang et al.~\cite{jiang2017mentornet} proposed the MentorNet to train an extra network with \ff{a clean validation set to select clean instances from the entire training database.}
When the clean validation data is unavailable, MentorNet adopted a self-paced curriculum to figure out reliable samples for training.
In another aspect, Han et al.~\cite{han2018co} proposed co-teaching, which simultaneously trains two networks in a symmetric way to select small loss samples for training. 
However, in the co-teaching framework, only part of the training data can contribute to the learning process in each epoch, which leads to unwanted waste of the valuable data samples with ``noisy" annotations.
In practice, medical images usually contain hard samples that also have large training loss due to different image quality/imbalanced class distribution. Discarding these hard samples will further change the class distribution and damage the representative ability of the learned deep features.

%
In this paper, we aim to develop a robust medical image classification method utilizing the noisy labeled data inherent in the training procedure. 
We propose a novel co-training framework with global and local representation learning for effectively tackling the sample selection bias and adopting all of the training data without wasting it.
We simultaneously train two student-teacher networks and each student-teacher network learns from each other to alleviate the bias caused by the noisy label. 
In particular, in each training cycle, the two student-teacher networks are trained independently for one epoch, followed by a noisy label filter (NLF) to automatically divide the training data into clean samples and noisy samples (or hard samples) according to the predictions of the teacher encoder.
Then the divided samples are crossly fed into the peer student-teacher network as input training data.
During training, the student-teacher network directly utilizes the clean samples in a fully supervised learning manner such that those noisy labels make less influence on model optimization. 
More importantly, we design a novel global and local representation learning scheme to employ the noisy samples in a self-supervised manner. This scheme includes a local contrastive loss to regularize the sample local clustering by pulling the similar samples together and pushing different samples far away; and a global \ff{relationship} consistency loss, applied on all the samples, to regularize the sample structure relationship consistency.
In the testing phase, the final prediction result is acquired by averaging the prediction of the two teacher models. 

Our main contributions can be summarized as follows:
\begin{itemize}
    \item We propose a novel self-supervised loss with global and local regularization to preserve the sample utilization efficiency and alleviate the overfitting of noisy labels.
   \item We develop a new self-ensemble co-training framework for robust medical image classification. We employ a student-teacher network instead of a single network to more robustly filter the clean samples and thereby eliminate the disturbance from noisy samples.
   \item The presented noisy label learning strategy has been extensively validated on three typical medical image noisy labels with public datasets, \ie, random noise, inter-rater variability, and computer generated noise. The proposed method consistently improves the prediction accuracy across different datasets and outperforms other learning from noisy labeled data methods.
\end{itemize}

%% file: 2-relatedwork.tex
\section{Related Work}
\label{sec:noisy}
\subsection{Deep Learning with Noisy Labels}
Although some approaches have been considered to address the noisy label issue, it is still an ongoing issue in the deep learning community. 
Existing literature on training with noisy labels focuses primarily on two directions. (1) estimate the transition matrix approaches  (\cite{hendrycks2018using,jiang2018hyperspectral,xia2019anchor,yao2020dual}). 
Sukhbaatar et al.~\cite{sukhbaatar2014training} proposed a noise transformation to estimate the transition matrix, which needed to be periodically updated and is non-trivial to learn. Patrini et al.~\cite{patrini2017making} designed a two-stage framework to learn on the noisy labeled training data, where in the first stage the transition matrix was estimated, and a correction loss was employed in the second stage. Similarly, Goldberger et al.~\cite{goldberger2016training} proposed to estimate the latent label transition matrix using deep learning method in an end-to-end manner by adding an extra layer.  
The accuracy of the classifier trained on noisy labeled data can be improved by such accurate estimation. However, the label transition matrix is hard to be estimated in reality due to complicated data biases. \ff{Moreover, the label noises in the medical image are usually instance dependent noise, which makes the estimation of transition matrix even challenging.}
(2) Sample selection strategy. Some works are designed to re-label/weight selected instances~\cite{tanaka2018joint,ren2018learning} or train on selected clean instances~\cite{chen2019understanding,jiang2017mentornet,han2018co,nguyen2019self,li2020dividemix}.
Tanaka et al.~\cite{tanaka2018joint} proposed a joint optimization framework, where the label of training data was updated during the training process. 
Jiang et al.~\cite{jiang2017mentornet} proposed to select clean instances for network training according to extra clean data. However, the extra clean data is not practical in clinical. They adopted a self-paced curriculum to figure out reliable samples for training when the extra clean validation data is unavailable.
Han et al.~\cite{han2018co} presented a co-teaching learning paradigm by simultaneously training two models and removing the potential noisy samples in each mini-batch according to the training loss of each input data. 
Besides, some recent works also adopted a semi-supervised learning strategy to address the noisy label problem  (\cite{ding2018semi,nguyen2019robust,nguyen2019self,xue2020cascaded,liu2020earlylearning}) by generating pseudo labels or adding regularization. Our method uses a self-supervised \ff{strategy} to effectively utilize the samples by specifically designed inter-patient relationship loss and local contrastive loss.

\begin{figure*}[t]
    \centering
    \includegraphics[width=1.0\textwidth]{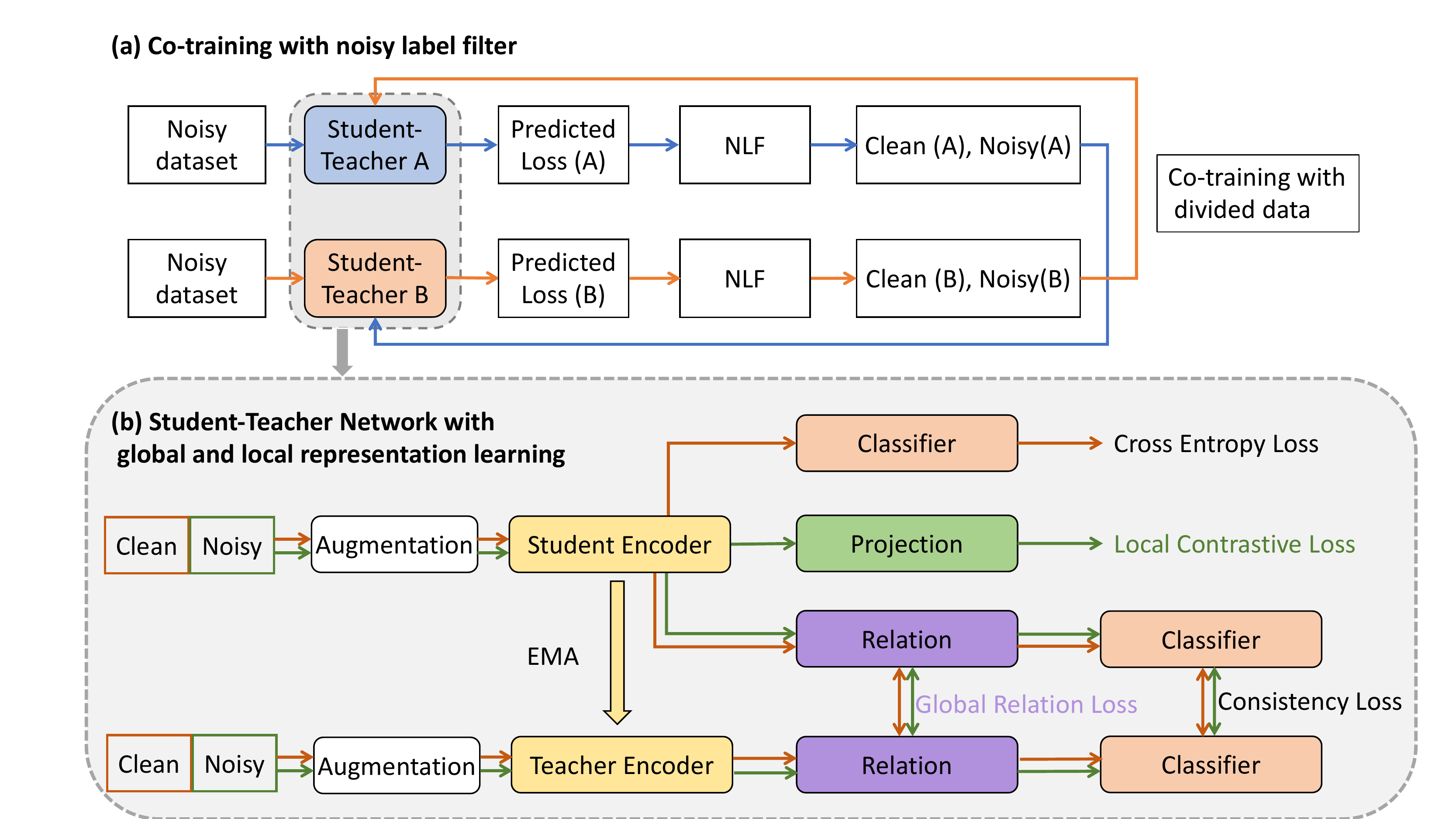}
    \caption{An overview of the proposed co-training with global and local representation learning framework: (a) is the co-training scheme with two \ff{independent} student-teacher networks and a noisy label detection procedure. NLF is the noise label filter. (b) represents the detailed training procedure of the teacher-student network. The network directly employs the clean samples with supervised cross-entropy loss. The local contrastive loss is applied on noisy samples and the global relation loss is applied on both clean and noisy samples. The weights of the teacher encoder are updated by the exponential moving average (EMA) of the student encoder.}
    \label{fig:flow}
\vspace{-10pt}
\end{figure*}
\subsection{Noisy Label Learning for Medical Image Analysis}
The annotation quality of medical images is prone to experience, which requires years of professional training and domain knowledge. 
\xcc{There exist many ambiguous images that will confuse the clinical experts, and therefore, results in misdiagnosis/wrong annotation, and disagreements.} 
\xc{For some large-scale datasets, the annotation is also heavily relied on automatically extracting labeling from radiological reports by natural language processing tools, which inevitably results in a certain level of label noise.}
Manually reducing those incorrect annotations requires agreements between experts and is time-consuming. Some studies have addressed the existence of noisy labeled medical images during the network training process.  Dgani et al.~\cite{dgani2018training} proposed to estimate the transition matrix of the noisy label during the training process by optimizing an extra softmax layer, it was heavily dependent on the noise label transition matrix assumption. Shu et al.~\cite{shu2019lvc} proposed to weaken the influence of noisy annotation in segmentation tasks by utilizing potential visual guidance during the learning process. To suppress the influence of noisy labels, some methods proposed a sample selection strategy to train deep learning models on selected data or assign appropriate weight to \xcc{the} training data. Xue et al.~\cite{xue2019robust} proposed a robust learning framework for noisy labeled medical image classification, where the severe hard samples and noisy labels in medical images are both considered by a sample selection module and a sample re-weighting strategy. Min et al.~\cite{min2019two} presented a semi-supervised biomedical segmentation network with noisy labels, where the noisy label issue was tackled by weakening the noisy gradients in multiple layers. Zhu et al.~\cite{zhu2019pick} proposed to evaluate the quality of the labels and use good ones to turn the network parameters. Le et al.~\cite{le2019pancreatic} utilized a small set of clean training samples and assigned weights to training samples to deal with sample noise. However, the previously proposed sample selection strategy has not considered the selection bias caused by the erroneous guiding of noisy samples, and only part of the valuable training data is used. 

The main challenge for medical image analysis with the noisy label has two aspects: (1) the difficulty of noisy label detection. Many hard samples (Fig.~\ref{fig:intro}) are inherently hard to learn and can be confused with noisy ones (\cite{xue2019robust,wang2018iterative}). However, directly discarding detected noisy samples will result in the missing of hard samples that are essential for discriminative feature learning. 
(2) How to utilize the detected noisy labeled data. Some studies discarded samples with a high probability of being incorrectly labeled or maintain these samples in a semi-supervised manner by refining those labels (\cite{li2020dividemix,ding2018semi}). 
The semi-supervised learning would be an ideal approach when the training data has a large proportion of noisy \xc{labels}. \xcc{However, the noisy label proportion of medical images is lower than nature images acquired from the web.
For a dataset with fewer noisy labels, the generation of the \ff{pseudo labels} will introduce more noise to the labels, especially for those hard samples. }

\subsection{Skin Lesion and Lymph Node Analysis}

Automated classification of skin lesion images is a challenging task owing to the fine-grained variability in the appearance of skin lesions. 
Deep convolutional neural networks show potential for general and highly variable tasks across many fine-grained object categories. A few deep learning algorithms have been \ff{proposed for} dermoscopy images~(\cite{codella2018skin,demyanov2016classification,kawahara2016deep,premaladha2016novel,yu2016automated}). \xcc{Premaladha et al. \cite{premaladha2016novel} adopted deep learning and hybrid AdaBoost Support Vector Machine (SVM) algorithms to classify melanoma skin lesions. This system was tested on 992 images and obtained a high classification accuracy (93$\%$)}. Kawahara et al.~\cite{kawahara2016deep} employed a fully convolutional network to extract multi-scale features for melanoma recognition, the author reported accuracy of 85.8$\%$ on a \ff{five-class classification} problem. Yu et al.~\cite{yu2016automated} applied a very deep residual network with a two-stage framework to distinguish melanoma from non-melanoma lesions.

A series of studies related to whole-slide image analysis \ff{have} been presented for a variety of classification, detection, and segmentation tasks. 
Wang et al.~\cite{wang2016deep} detected the lymph node in whole slice images with the ensemble of two GoogLeNets. Kong et al.~\cite{kong2017cancer} detected cancer metastasis using a spatially structured deep network with appearance and spatial dependencies. Courtio et al.~\cite{courtiol2018classification} detected the lymph node metastases in a weakly supervised manner with feature embedding and multiple instance learning techniques. Liu et al.~\cite{liu2017detecting} adopted Inception architecture with patch sampling and data augmentations to aid breast cancer metastasis detection in lymph nodes. Cruz et al.~\cite{cruz2017accurate} presented a classification approach for detecting the presence and extent of invasive breast cancer on the whole slide digitized pathology images using a CNN classifier.

%% file: 3-method.tex
\section{Methods}
\subsection{Problem Setting}
In supervised learning, we consider finding a mapping function $f\!: \mathcal{X} \rightarrow \mathcal{Y}$, where $\mathcal{X}$ is image space, $\mathcal{Y}$ is label space, and $f$ describes the complex relationship between them. To learn the mapping function $f(\cdot)$, a loss objective is usually defined to penalize the observed differences between the model prediction $f(x)$ and \ff{the} ground truth target $y$ for a training sample pair. The aim is to learn a classifier that can predict a label for the given observation $x$ at testing time. 
Typically, an estimated risk $R(f)$ is computed with the observed set of training samples
\begin{equation}
    R(f)=\mathbb{E}_{(x,y)\thicksim P_{\mathcal{X},\mathcal{Y}}}[\mathcal{L}(f(x),y)].
\end{equation}
The optimal classifier is the one which minimizes the expected risk,\ie, $f^\ast=\textit{argmin}\ R(f)$.

More specifically, in our problem setting, we denote $x$ as an image sample and $\hat{y}$ as its observed label but possibly noisy (\ie, incorrect). The real ground truth label $y$ for the sample is unknown due to various annotation limitations, misdiagnosis, or disagreements. During training, we only have $\hat{y}$ assigned to each sample, \ff{which means that for a noisy labeled training dataset, there are input images $x$, the noisy labels $\hat{y}$, and the latent clean labels $y$.}
In our approach, we train a network based on $\{x,\hat{y}\}$ with underlying $\{x,y\}$ unavailable. It has been reported that naive training with the noisy labels $\hat{y}$ will result in performance degradation for predictions on test set~(\cite{zhang2016understanding,chen2019understanding}). With our proposed robust learning strategy, we aim to optimize the CNN classifier with $\{x,\hat{y}\}$ while achieving comparable results with a model trained on $\{x,y\}$.

\begin{figure}[t]
\begin{tabular}{cc}
  \includegraphics[width=0.47\columnwidth]{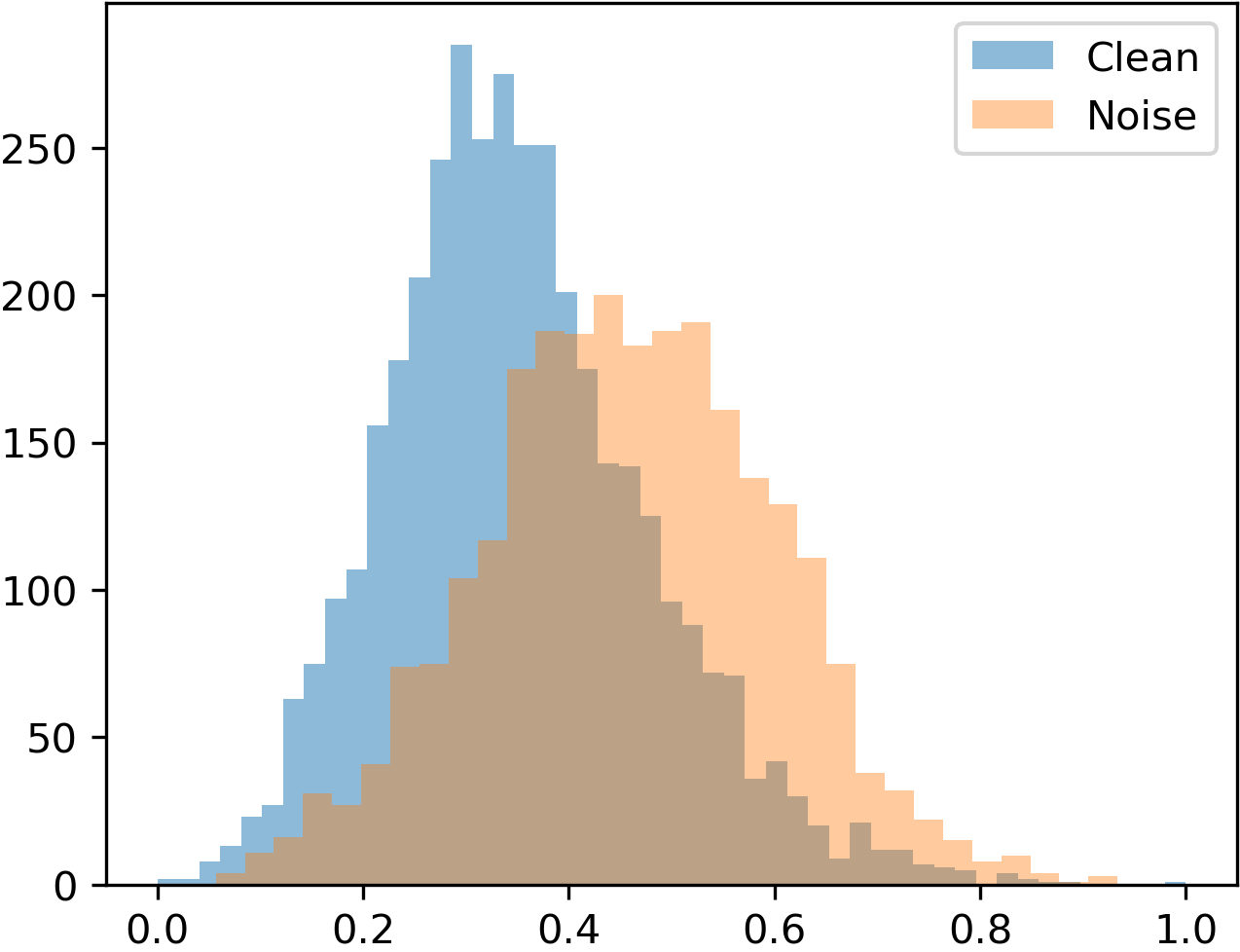} & \hspace{-5mm} \includegraphics[width=0.47\columnwidth]{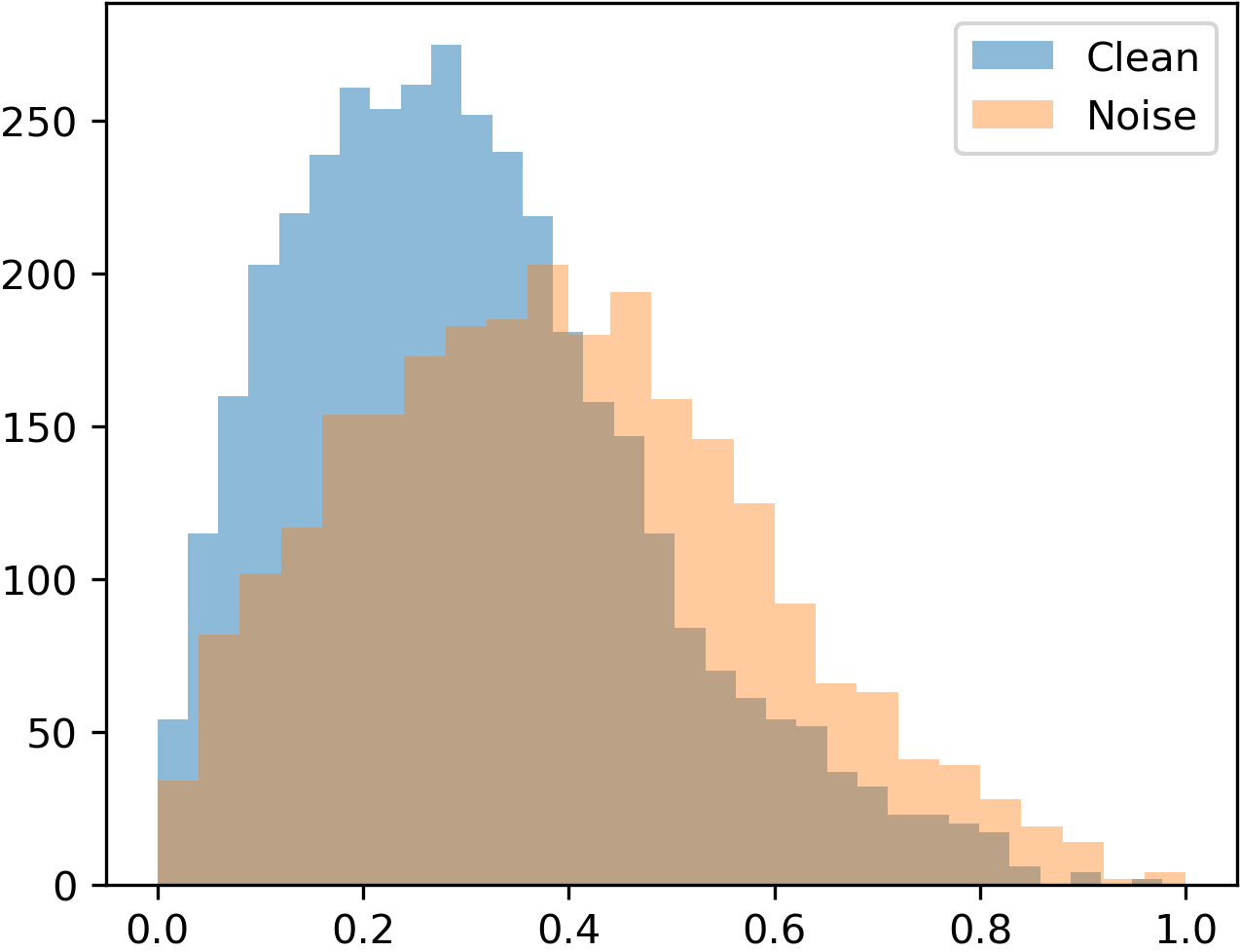}  \\
(a) &\hspace{-10mm} (b) \\[2pt]
\end{tabular}
\caption{The per-sample training loss of the lymph node histpathologic images with 40$\%$ random noise trained by cross entropy. (a) The per-sample loss distribution after 10 epoch, the different colors represent different classes. (b) The per-sample loss distribution after 20 epoch. We plot the distribution of clean samples and noise samples. }
\label{fig:loss}
\vspace{-8pt}
\end{figure}

\subsection{Co-training Paradigm with Noise Label Filter}
To alleviate the error flow coming from the biased selection of \ff{the} training instances, we adopt the two networks training scheme as~\cite{han2018co,malach2017decoupling,chen2019understanding}. 
Fig.~\ref{fig:flow} elaborates our proposed self-ensemble co-training paradigm.
Our framework maintains two independent teacher-student networks trained simultaneously, denoted as \textit{student-teacher A} and \textit{student-teacher B}. These two networks have different weight parameters due to the different initialization, different input image augmentation, different training data division, and different mini-batch sequences so that they can filter out the noisy labels independently.
Fig.~\ref{fig:flow}(a) shows the procedure of one training cycle, after train the student-teacher network for one epoch, all the training data go through a noisy label detection process, where the \textit{predicted loss (A)} and \textit{predicted loss (B)} of all the samples on \textit{student-teacher A} and \textit{student-teacher B} are calculated. Then we fit the \textit{predicted loss (A)} in a noisy label filter to acquire a clean group and a noisy group, denoted as clean (A), noisy (A). The same process for \textit{predicted loss (B)} to acquire clean (B), noisy (B). The divided clean and noisy samples are crossly sent to \ff{the peer} student-teacher network as new input data. For example, we train \textit{student-teacher network A} by (clean (B), Noisy (B)) and vice versa. With the renewed dataset from its peer network, each student-teacher network trained independently in the next cycle. We use the blue and orange colored box and arrow to show the data flow.
Different from Co-teaching~\cite{han2018co} that selected small loss samples and proceed crossed training in each mini-batch, we perform crossed training after each epoch with the divided training samples.

To tackle the challenge of noisy label detection, after each epoch we predict the cross entropy loss of all the training data by two teacher networks respectively and input the cross entropy loss into a noisy label filter (NLF) separately as shown in Fig. \ref{fig:flow} (a). \ff{The NLF} adopts a two-component Gaussian Mixture Model to fit the max-normalized loss of the training data using the Expectation-Maximization algorithm in which the clean and noisy components are divided in an unsupervised manner. For each sample $l_i$, its clean probability $w_i$ is the posterior probability $p(g|l_i)$, where $g$ is the Gaussian component with smaller mean. The noisy label filter divides the training samples into clean set and noise set by setting a threshold $t$ on $w_i$. During training, we employ a linear schedule $t$, that $t$ gradually decreased from $0.9$ to $0.5$ in the first 10 epochs, and maintain $0.5$ in the following epochs. The strategy can select the most confident samples at the initial stage, then gradually reduce the confidence requirement as the model is becoming more robust. 
The initial threshold is $0.9$ for the histology lymph node images since it is a balanced dataset, \ff{while for the other dataset}, we set the initial threshold to $0.8$ to avoid neglect of less represented class.

To improve the performance of noisy label detection, we propose to adopt a student-teacher network for better feature learning.
Fig.~\ref{fig:flow}(b) shows the training scheme of the student-teacher network, which adopts the augmented clean and augmented noisy dataset as input for the student encoder and teacher encoder. The whole network is trained by four losses that will be introduced in the next section. The student-teacher network is proposed for semi-supervised learning, while we adopt it in the noisy label learning scenario. As we only update the student network with the selected small loss samples and update the weights of the teacher model by the exponential moving average (EMA) of student network weights.
The teacher model is more robust to noisy labels and also can enhance the separation of hard samples and noisy labeled data implicitly. 

Since we observe that the deep learning model always first learns the easy instances and adapts to \ff{hard} noisy instances, as the separable of clean and noisy data gradually decreased as shown in Fig.~\ref{fig:loss}. We design a warm-up stage with five epochs, in which the two networks are updated by all the samples.


\subsection{Self-supervised Learning for Noisy Samples}
\label{sec:representation}
After the noisy label detection procedure, instead of discarding the noisy labeled sample, we propose to utilize the noisy labeled samples by the self-supervised learning strategy, which can mitigate the requirement for labeled data by providing a means of leveraging unlabeled data.
Some previous studies \ff{generated} refined labels for selected noisy labeled samples, which \ff{performed} well on the heavily corrupted dataset. However, severe noise is not common in medical image scenarios. A refined label would introduce more noise to the training label in the mild noisy situation. 
To learn better feature representation, we propose to explicitly regularize the unlabeled data in the feature space. The regularization is designed in two perspectives: the global domain, where we align the per-sample relation matrix of each batch to preserve the general knowledge of the inter-sample relationships; and the local domain, where we add a contrastive loss to the noisy labeled samples~\cite{chen2020simple} to strength the feature representation learning. 
Thus, both the noisy sample and hard samples can contribute to \ff{the} feature learning.

\para{Global inter-sample relationship alignment.}
The relation among samples widely exists in medical imaging. In clinical diagnosis, experienced clinicians tend to make a diagnosis according to the reference from previous analogous cases~\cite{avni2010x}. For each batch of noisy data, the prediction of student network and teacher network should maintain the same relationships. We calculate an inter-sample relation matrix that can capture the relationships between each sample within a mini-batch. 
Specifically, for each mini-batch with B samples, we extract the feature before the $fc$ layer of the backbone network, the shape of the feature map in ResNet34 is $F \in R^{B \times 512}$. The relation matrix is computed as:
\begin{equation}
\label{eq:relation}
    G=F \cdot F^T
\end{equation}
where $G_{ij}$ is the inner product between the vectorized activation map $F_{(i)}$ and $F_{(j)}$, whose intuitive meaning is the similarity between the $i_{th}$ sample and $j_{th}$ sample within the input mini-batch. The final sample relation matrix $M$ is obtained by conducting the L2 normalization for each row $G_i$ of $G$. We then propose a global inter-sample relationship loss $L_{global}$ to align the inter-sample relationship of samples between the teacher model and student model by minimizing their symmetrized Kullback–Leibler (KL) divergence

\begin{equation}
\label{eq:kl}
    L_{global}= \frac{1}{2} (D_{KL}(M^{s}||M^{t})+D_{KL}(M^{t}||M^{s})),
\end{equation}
where $D_{KL}(p||q)=\sum_i p_i log \frac{p_i}{q_i}$, $M^{s}$ and $M^{t}$ represent the sample relation matrix $M$ of student network and teacher network respectively. This global inter-sample relationship loss is applied to all the samples.

\para{Local contrastive loss.}
In addition to promoting the alignment of feature relationships with $L_{global}$, we further encourage robust feature representation by adding a local self-supervised contrastive loss following \cite{chen2020simple}, denoted as $L_{local}$. \ff{This loss considers the clustering of sample features regardless of the label}
Intuitively, the $L_{local}$ learns representations by maximizing agreement between two augmented images of the same data in a latent space.

Specifically, we replace the $fc$ layer of the backbone network by a projection layer ($g(\cdot)$) to embed the feature to 128 dimensions (denoted as $z_i$). The projection layer includes one hidden layer and ReLU. The local contrastive loss is calculated for each sample in the mini-batch. We augment each image to two images, so that the difference between positive pairs (two augmented images from same source image) can be minimized while the difference from negative pairs (augmented images from different source) is maximized. Let $i \in {1...2N}$ be the index of an arbitrary augmented image, and let $j(i)$ be the index of the other augmented image originating from the same source image. The local contrastive loss is 
\begin{equation}
\label{eq:contrastive}
\begin{aligned}
    & L_{local}=\sum_{i=1}^{2N} L_i^{local}\\
   & L_i^{local}=-log \frac{exp(sim(z_i , z_{j(i)})/\tau)}{\sum_{k=1}^{2N} 1_{i \neq k} \cdot exp(sim(z_i, z_k) / \tau)}\\
\end{aligned}
\end{equation}
where $z_i$ and $z_j$ denote the projection of features, $\tau$ is a scalar temperature parameter. We analyze the influence of $\tau$ in the experiment section. $sim(z_i,z_j)=z_i^T z_j/||z_i|| ||z_j||$ denote the cosine similarity between $z_i$ and $z_j$.

\subsection{Overall Loss Functions}
Given the batches in each iteration, the total loss consists of cross-entropy loss $L_{CE}$ between the observed labels and model predictions for the clean dataset and the unsupervised loss, which includes the original consistency loss $L_{con}$ of mean-teacher for both clean and noisy labeled dataset, the local contrastive loss $L_{local}$ for noisy labeled data, and global inter-sample relationship alignment loss $L_{global}$ for both clean and noisy labeled data. 
Specifically, the loss function for clean data is the cross entropy loss
\begin{equation}
\label{eq2}
    L_{CE} = -\frac{1}{N_{clean}} \sum_{i \in N_{clean}} \sum_{j \in C}y_i^j \log f_s^j (x_i,\theta),
\end{equation}
where $N_{clean}$ is the number of clean samples and $C$ is the number of class.
Also, the original consistency loss is adopted 
\begin{equation}
\label{eq:con}
    L_{con} = \frac{1}{N} \sum_{i \in N} ||f_{s}(x_{i},\theta_{s})-f_{t}(x_{i}',\theta_{t}))||_2^2,
\end{equation}
where $x$ and $x'$ represent two differently augmented input images, $\theta_{s}$ and $\theta_{t}$ are the weights of student model and teacher model, N is the total image number.
Then the total objective function can be represented as
\begin{equation}
\label{eq:total}
L_{total}  =L_{CE} + \lambda(L_{global}+L_{local}+L_{con}),
\end{equation}
where $\lambda$ is the weight of unsupervised loss. Follow the common practice (\cite{tarvainen2017mean,berthelot2019mixmatch}), we linearly ramp up $\lambda$ from 0 to 10 over the first 10 epochs after warm-up procedure. 
More details on the effect of $\lambda$ and each loss item are shown in the experiments.


\subsection{Implementation details}
We trained two mean-teacher networks simultaneously with the data selected by their peer network. The exponential moving average decay (EMA) of the mean-teacher network was 0.99 according to \cite{tarvainen2017mean}. The temperature of local loss $\tau$ was 0.5, and unsupervised loss weight $\lambda$ was 10. We adopted the ImageNet pre-trained ResNet34 with randomly initialized fully connected layer as the backbone for the skin lesion, lymph node, and gleason grading datasets, DenseNet121 as the backbone for NIH dataset. The network was trained with Adam optimizer, with weight decay as 1e-4, and an initial learning rate of 1e-3. We decreased the learning rate after each epoch by 0.9. We trained the network using one NVIDIA TITAN Xp GPU and the batch size was set to 64. We adopt random resize and crop with a scale of 0.2 to 1, random rotation, random horizontal flipping as augmentation methods.

%% file: 4-experiment.tex
\section{Experiments}
\xcc{To evaluate the effectiveness of our proposed self-ensemble co-training framework for noisy label learning, we have conducted extensive experiments on four challenging medical image diagnosis tasks: 1) melanoma lesion classification from dermoscopy images, 2) lymph node classification from the histopathologic images, 3) Chest X-ray classification \ff{with} NLP labeled datasets, and 4) prostate cancer gleason grading using the digital histopathology images. Comparison with a set of current state-of-the-art noisy label training methods and insightful analytical studies will be elaborated in this section.}
\begin{table*}[t!]
\centering
\caption{Comparison of classification results with state-of-the-art methods on lymph node data (Accuracy, \%).}
\label{tab:hist-comp}
 \resizebox{0.9\linewidth}{!}{
\begin{tabular}{c|c|c|c|cc|c|c}
\toprule[1.5pt]
 Noise ratio & Cross entropy     & F-correction (2017)   & Mentornet (2018)  & Decoupling (2017) &Co-teaching (2018)  & ELR (2020) & Ours        \\ \hline
0   &93.44 $\pm$ 0.41   &94.22 $\pm$ 0.19 &93.09 $\pm$ 0.21  &93.03 $\pm$ 0.35    & 93.25 $\pm$ 0.28  & 93.27 $\pm$ 0.20 & \textbf{94.31 $\pm$ 0.14}  \\ 

 0.05       & 91.28 $\pm$ 0.57  &92.60 $\pm$ 0.35   &92.65 $\pm$ 0.32      &91.73 $\pm$ 0.47    &  92.88 $\pm$ 0.36  &  92.69 $\pm$ 0.33     & \textbf{93.88 $\pm$ 0.35} \\
0.1        & 88.95 $\pm$ 0.31 & 90.54 $\pm$ 0.18 & 91.05 $\pm$ 0.27     & 89.74 $\pm$ 0.43   &  91.39 $\pm$ 0.48 &  90.84 $\pm$ 0.24     & \textbf{92.75 $\pm$ 0.30}  \\
0.2        & 82.67 $\pm$ 0.49 & 83.24 $\pm$ 0.26      &85.66 $\pm$ 0.27     &  82.85 $\pm$ 0.38   &    86.05 $\pm$ 0.17  &    87.12 $\pm$ 0.28 & \textbf{90.88 $\pm$ 0.20}    \\
0.4        & 63.41 $\pm$ 0.63 & 64.10 $\pm$ 0.37     &69.43 $\pm$ 0.47      & 64.13 $\pm$ 0.24    &   76.03 $\pm$  0.18  &   73.86 $\pm$  0.17   & \textbf{79.00 $\pm$ 0.34}     \\ 
\bottomrule[1.5pt]

\end{tabular}
}
\end{table*}
\begin{table*}[t!]
\centering
\caption{Comparison of classification results with state-of-the-art methods on ISIC melanoma data (Accuracy, \%).}
\label{tab:skin_compare}
 \resizebox{0.9\linewidth}{!}{
\begin{tabular}{c|c|c|c|cc|c|c}
\toprule[1.5pt]
 Noise ratio & Cross entropy     & F-correction (2017)   &Mentornet (2018)  & Decoupling (2017) &Co-teaching (2018)  &ELR (2020) & Ours        \\ \hline
0   &85.50 $\pm$ 0.28   &85.63 $\pm$ 0.14 &85.09 $\pm$ 0.21  &84.98 $\pm$ 0.14  &85.59 $\pm$ 0.27 &85.51 $\pm$ 0.21 &\textbf{86.00 $\pm$ 0.12}  \\
           
0.05       & 84.25 $\pm$ 0.66 &84.32 $\pm$ 0.39        &83.47 $\pm$ 0.27      &83.39 $\pm$ 0.22   & 83.98 $\pm$ 0.45  & 84.28 $\pm$ 0.23& \textbf{85.40 $\pm$ 0.22}\\

0.1        & 83.01 $\pm$ 0.37 &83.82 $\pm$ 0.46       & 83.83 $\pm$ 0.37    & 83.13 $\pm$ 0.44   &  83.39 $\pm$ 0.50  &  83.18 $\pm$ 0.35     & \textbf{84.33 $\pm$ 0.31} \\
0.2        & 81.36 $\pm$ 0.61 &82.88 $\pm$ 0.61        &82.15 $\pm$ 0.49      & 81.64 $\pm$ 0.53    &  83.22 $\pm$ 0.21    &  83.37 $\pm$ 0.48     & \textbf{84.17 $\pm$ 0.32} \\
0.4        & 69.65 $\pm$ 0.65 & 68.60 $\pm$ 0.50        & 70.40 $\pm$ 0.33     & 68.63 $\pm$ 0.47   &  74.78 $\pm$ 0.44  &  72.98 $\pm$ 0.26     & \textbf{76.67 $\pm$ 0.14 }\\ 
\bottomrule[1.5pt]
\end{tabular}}
\end{table*}
\subsection{Dataset and Pre-processing}

The ISIC 2017 skin dataset comes from the melanoma competition held in conjunction with ISBI 2017. This is a binary classification task \ff{with} malignant and benign \ff{class}. The dataset consists of 2000 melanoma dermoscopy images for training and 600 melanoma dermoscopy images for testing. Besides these data, we also downloaded extra 1582 images from the ISIC archive for training. We resized all the skin images to the size of 224 $\times$ 224 and normalized each image by subtracting ImageNet mean and std.

The Kaggle histopathologic cancer detection dataset is a slightly modified version of the PatchCamelyon (PCam) \cite{bejnordi2017diagnostic,veeling2018rotation}, which removed the duplicate images. It packs the clinically relevant task of metastasis detection into a straight-forward binary image classification task into malignant and benign. A positive label indicates that the center 32 $\times$ 32 region of a patch contains at least one pixel of tumor tissue. Tumor tissue in the outer region of the patch does not influence the label. The dataset contains 220025 images for training, 57460 images for online testing. Without loss of generality, we randomly selected 6200 and 800 images from the training images as our training data and test data, respectively. We resized all the images to the size of 224 $\times$ 224, and normalized by subtracting ImageNet mean and std. 

\xcc{Besides the two datasets mentioned above, we also tested our method on two applications with real noise.} The first dataset is the NIH chest X-ray dataset \cite{wang2017chestx}, which contains 112,120 frontal-view CXR images from 32,717 patients. Each image is labeled with 14 possible pathological findings that are automatically mined from the text reports. We used 1,962 manual labeled images \cite{majkowska2020chest} as clean testing data. We resized all the skin images to the size of 224 $\times$ 224 and normalized each image by subtracting ImageNet mean and std.

\xcc{The second dataset is the Gleason 2019 dataset that contains 333 tissue microarray (TMA) images of prostate cancer, which are sampled from 231 radical prostatectomy patients. All the images are annotated in detail by six pathologists for different gleason grades. The task is to classify the score into benign (0) and cancerous (1,2,3,4,5). The data collection and labeling have been introduced in \cite{nir2018automatic}. This dataset is annotated by six pathologists with different experience levels, hence it contains images with high inter-observer variability. We resized the image to the size of 224 $\times$ 224 and normalized each image by subtracting ImageNet mean and std.}

\subsection{Experimental Settings}
\label{sec:data}
\para{Random noise.}
\xcc{We first \ff{evaluated} our method on two synthetic noisy label datasets. To simulate the noisy label situations on our employed melanoma dataset and lymph node dataset}, we randomly sampled $\gamma \in \{0.05, 0.1 ,0.2, 0.4\}$ percentage of images from each class and flip the labels of these images following the common setting. In this way, we can obtain a noisy training dataset with a certain amount of images having incorrect labels.
The noisy label is defined as $y^{\prime}_{i}=y_{i} $ with the probability of $ 1-\gamma$, and $y^{\prime}_{i}=y_{k},y_{k}\neq y_{i} $ with the probability of $\gamma$, where $y^{\prime}_{i}$ is the corrupted noisy label, $y_{i}$ is clean label. We \ff{adopted} the accuracy as evaluation metric.
\begin{figure}[t]
    \centering
    \includegraphics[width=0.47\textwidth]{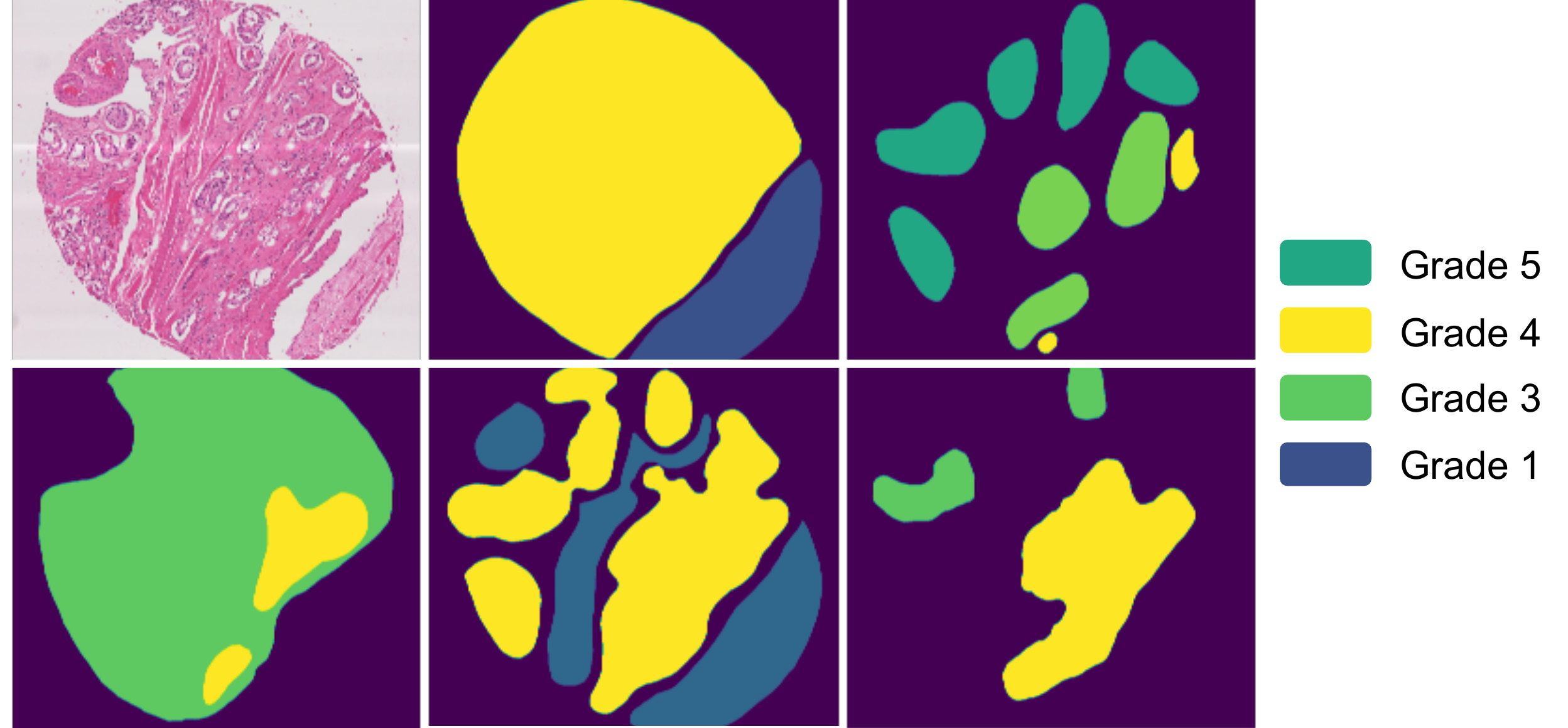}
    \caption{The Gleason grading from different pathologists for one sample histological image.}
    \label{fig:gle}
\vspace{-5pt}
\end{figure}
\para{Computer-generated noisy label.}
\xcc{We trained our method on the NIH Chest X-ray dataset to simulate the reality and then further tested our trained model on a manually relabeled chest X-ray dataset to evaluate the effectiveness of our method. This dataset has four labels: pneumothorax, opacity, nodule or mass, and fracture. Only pneumothorax, nodule or mass are labeled in the NIH dataset. To be consistent with the annotations used in the paper, we \ff{averaged} the outputs of Mass and Nodule to be the prediction of Mass or Nodule.}

\para{Inter-observer variability.}
\xcc{The Gleason 2012 challenge dataset is labeled by six pathologists with an inter-observer Cohen's kappa coefficient between 0.4 and 0.6 (0.0 indicates chance agreement and 1.0 means perfect agreement) \cite{nir2018automatic,karimi2019deep}, as shown in Fig 4. Following \cite{karimi2019deep,karimi2020deep}, we applied the Simultaneous Truth and Performance Level Estimation (STAPLE) algorithm \cite{warfield2004simultaneous}, which is \ff{based on the expectation-maximization method to estimate the ground-truth labels from the annotations of six radiologists.} We utilized the STAPLE estimated labels as clean labels, and the label of one of the six pathologists as \ff{the} noisy labels. We used a patch size of $750 \times 750$, and the grading of the center $250 \times 250$ region as the label. For the experiment, we used 80$\%$ data for training and the rest for testing.}

\subsection{Comparison with state-of-the-art methods}

To show the effectiveness of our method, we \ff{conducted} a comparison with the state-of-the-art noisy label learning methods. The compared methods include:
\begin{itemize}
   \item Co-teaching \cite{han2018co}: They adopted a two network learning strategy and select the small loss sample for \ff{the training of the }peer network.
   \item Decoupling~\cite{malach2017decoupling}: They used a two network setting, and update the two networks based on the disagreement of each network.
   \item F-correction~\cite{patrini2017making}: They corrected the prediction by a noise transition matrix. As suggested by the authors, we first trained a standard network to estimate the transition matrix.
   \item MentorNet~\cite{jiang2017mentornet}: They used a group of clean data to weight each training data. We used self-paced MentorNet in this paper as we don't have the extra clean data.
   \item ELR~\cite{liu2020earlylearning}: They leveraged semi-supervised learning techniques and regularization term to prevent memorization of false labels. 
\end{itemize}
We evaluated the classification performance of these methods and our method by calculating their average classification accuracy or area under curve (AUC). The results are listed in Table~\ref{tab:hist-comp}, Table~\ref{tab:skin_compare}, Table~\ref{tab:nih}, and Table~\ref{tab:gleason} for the four datasets with three noise types.
For fair comparisons, we obtained the classification results of \ff{the} comparison methods by downloading their public implementations or reimplementing the method according to their paper. 
\xcc{Similarly, the ResNet-34 was employed as the network backbone for the ISIC skin cancer data, kaggle lymph node data, and the Gleason 2019 data. The Densenet121 was the backbone for the NIH chest X-ray data. Same hyper-parameters (learning rate, weight decay, optimizer) for training were adopted.} We applied the same strategy as~\cite{han2018co} to report the evaluation metrics of the last ten epochs and report the results averaged over three runs.
 
           


\begin{figure*}[t!]
    \centering
    \includegraphics[width=0.85\textwidth]{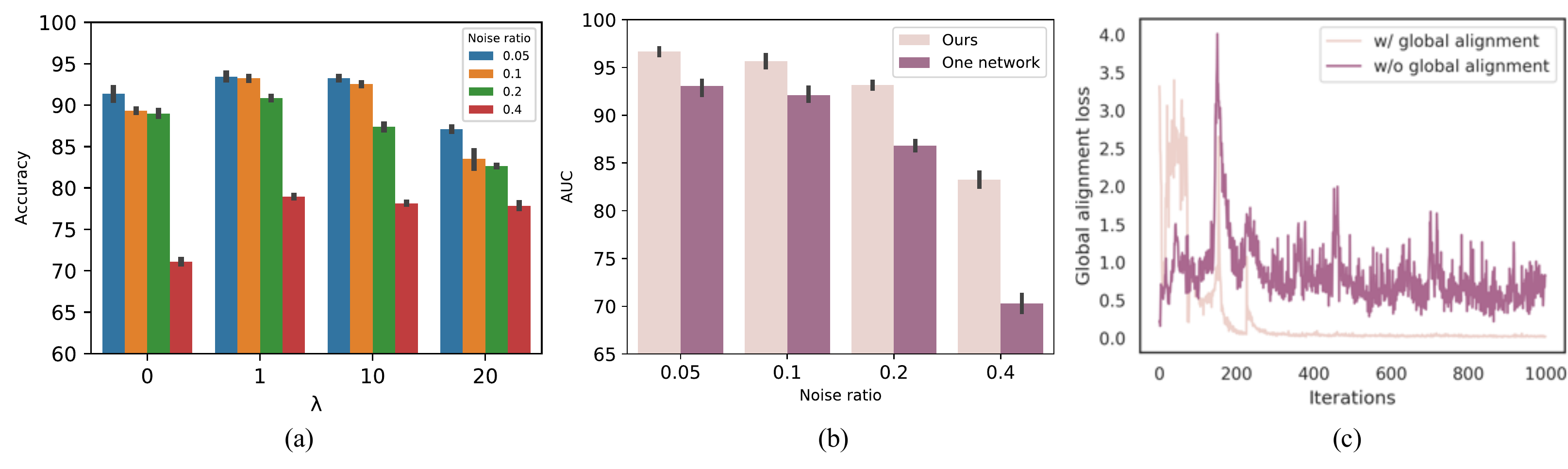}
    \vspace{-2pt}
    \caption{(a) The performance of our method under different unsupervised loss weights on the histopathologic dataset.; (b) The AUC value of selected clean training samples under different noise setting, compared between ours and one network setting without $L_{global}$ and $L_{local}$ on the lymph node dataset.; (c) The change of global inter-sample relationship during network training under different scenarios.}
    \label{fig:auc}
\end{figure*}
\begin{table}[t!]
\centering
\caption{\xc{Results on the NIH dataset (AUC).}}
\label{tab:nih}
\begin{tabular}{c|cc}
\toprule[1.5pt]
Method &Pneumothorax & Nodule or Mass \\ \hline
Cross entropy &0.870 &0.843  \\
F-correction & 0.808 & 0.848 \\
Mentornet    & 0.866      &0.837       \\
Decoupling   & 0.801   & 0.843 \\
Co-teaching  & 0.873 & 0.820  \\
ELR          & 0.871 & 0.832\\
Ours &0.891 &0.846  \\
\bottomrule[1.5pt]
\end{tabular}
\end{table}

\para{Results on histopathologic lymph node dataset.}
Table~\ref{tab:hist-comp} shows the results of our method and other competitors on the lymph node classification task. For the clean dataset (\ie, noise ratio=0), only F-correction and our method obtained better results than \ff{the} cross entropy loss. This phenomenon shows the drawback of the sample selection-based training strategy, \ff{where the hard samples are treated as the noisy label and removed}. 
For all the noise ratio settings, decoupling obtained less improvement over cross entropy loss, indicating that even with two network settings, the training strategy that updates the network according to \ff{disagreements} may not work well in the medical image domain. For the noisy label dataset (noise ratio $>$ 0), Both Co-teaching and our method outperformed other methods by a large margin, indicating that the sample selection bias is efficiently alleviated by the two-network co-training scheme. Furthermore, our method outperformed Co-teaching in all the noise \xc{settings}, showing that by adding the noisy label detection and two unsupervised loss can learn more robust feature representation. Our method also achieved \ff{better} results than ELR, showing that the two unsupervised losses can reduce the effects of the noisy label. For \xc{the small noise ratio setting}, the overfitting to noisy label is not as severe as the heavily corrupted dataset, but all the methods achieved better results than training with cross entropy\xc{, and our method has the best results.}
\begin{table}[t]
\centering
\caption{Results on the Gleason 2019 challenge dataset (Accuracy, $\%$)}
\label{tab:gleason}
\begin{tabular}{c|cc}
\toprule[1.5pt]
              & A       & B \\ \hline
Cross entropy & 78.52 & 66.47         \\
F-correction  & 76.75        & 68.92         \\
Mentornet     & 79.80         & 68.21         \\
Decoupling    & 80.66        & 69.63         \\
Co-teaching   & 79.58        & 71.93         \\
ELR           & 80.37        & 70.27         \\
Ours          & 81.76        & 75.47    \\ 
\bottomrule[1.5pt]
\end{tabular}
\end{table}

\para{Results on ISIC melanoma dataset.}
Note that the skin lesion classification data is more challenging than the lymph node classification task, as the data is very imbalanced and contains more hard samples. From Table~\ref{tab:skin_compare}, it is observed that when there is no noise in the training data, Decoupling and Mentornet obtained lower accuracy than cross entropy loss, which is reasonable as Decoupling only utilized partial data for training and Mentornet possibly assigned smaller weight for hard samples. Those filtered or down-weighted samples are crucial for discriminative feature learning. In 0.05 noise ratio setting, Co-teaching does not perform well in the mild case due to the low sample utilization rate and wrong identification of hard samples; while self-paced Mentornet did not perform well due to the sample selection bias by single network setting and the down-weighted hard samples. 
F-correction obtained marginal improvement on accuracy when the noise ratio is small, but the performance dropped dramatically when the noise ratio reached 0.4, as the transition matrix estimation is not robust in this setting.
\xcc{Compared with our methods, the Co-teaching entirely removed the selected noisy label which would result in a worse classification performance due to the missing of hard samples.} Results demonstrate that our method can effectively work on the challenging imbalanced dataset, as the co-training scheme with noisy label filter can filter out the noisy labels and the two self-supervised \xc{losses} maintain an effective utilization of hard samples (minority group).

\para{Results on public CXR datasets.}
\begin{table*}[t!]
\centering
	\caption{Ablation study on skin lesion and lymph node (Accuracy, $\%$)}
	\label{tab:ab_hist}
\begin{tabular}{llll|llll|llll}
\toprule[1.5pt]
\multirow{2}{*}{Co-training+NLF} & \multirow{2}{*}{Self-ensemble} & \multirow{2}{*}{$L_{global}$} & \multicolumn{1}{c|}{\multirow{2}{*}{$L_{local}$}} & \multicolumn{4}{c|}{lymph node}     & \multicolumn{4}{c}{skin lesion}      \\ \cline{5-12} 
                     &                       &                         & \multicolumn{1}{c|}{}                       & 0.05  & 0.1   & 0.2   & 0.4   & 0.05  & 0.1   & 0.2   & 0.4   \\ \hline
                    -&                    -   &                      -   &                                         -    & 91.28 & 88.95 & 82.67 & 63.41 & 84.25 & 83.01 & 81.36 & 69.65 \\
                 $\checkmark$    &  -                     &   -                      &     -                                        & 91.65 & 89.33 & 84.00 & 68.25 & 84.37 & 83.49 & 82.53 & 73.49 \\
                  $\checkmark$   &                $\checkmark$     & -                        &   -                                          & 91.92 & 90.03 & 86.12 & 71.11 & 84.66 & 83.71 & 82.79 & 74.26 \\ \hline
                   $\checkmark$  &                $\checkmark$  &                     $\checkmark$    &  -                                           & 93.02 & 91,93 & 88.62 & 77.92 & 85.22 & 84.18 & 83.83 & 75.20 \\
                  $\checkmark$   &                     $\checkmark$  &                      -   &                     $\checkmark$                        & 92.90 & 91.94 & 88.50 & 77.75 & 85.07 & 84.02 & 83.27 & 75.04 \\
                  $\checkmark$   &                    $\checkmark$   &                       $\checkmark$  &                                 $\checkmark$            & 93.88 & 92.75 & 90.88 & 79.00 & 85.40 & 84.33 & 84.17 & 76.67 \\ 
                  \bottomrule[1.5pt]
\end{tabular}
\end{table*}
\begin{table}[t!]
\centering
\caption{The influence of temperature scale in contrastive loss (Accuracy, \%).}
\label{tab:temperature}
\begin{tabular}{c|cccc}
\toprule[1.5pt]
$\tau$ &0.01 &0.1 & 0.5 &1 \\ \hline
Skin lesion &84.50 &84.17 & 84.33 & 85.17 \\
Lymph node &91.88 &91.25 & 92.75 & 92.25 \\
\bottomrule[1.5pt]
\end{tabular}
\end{table}
\xcc{Manual annotation of large-scale datasets is time-consuming and expensive, some public datasets are often extracted by NLP from radiological reports, which inevitably results in a certain level of label noise. Wang et al.~\cite{wang2017chestx} reported the text mining accuracy of NIH Chest X-ray dataset on 900 manual labeled images, which shows that the precision and recall for Nodule and Pneumonia is below $90\%$. \xcc{Recently, Majkowska et al. \cite{majkowska2020chest} have relabeled a subset (1,962 images) of NIH testing data with at least three radiologists per image. We trained our model on the NIH datasets with NLP extracted label and tested on this relabeled data. Table \ref{tab:nih} shows the AUC values of our method and other comparison methods. Some of the comparison methods performed worse than the baseline, either for Pneumothorax or Nodule/Mass, we attribute this to the small noise proportion and unbalanced sample distribution of the 14 diseases, which makes the transition matrix estimation and sample selection challenging. As our method integrated the noisy sample selection method and self-supervised learning strategy, it still outperformed the baseline and other methods.} Overall, this experiment demonstrates the robustness of the proposed method to label noise in the real clinical setting.}

\para{Results on public Gleason grading datasets.}
To evaluate our method on noisy data, we consider the grading by one pathologist as the noisy label and the label estimated by STAPLE from six pathologists as the clean label. 
We conducted two groups of studies, \textit{A}: all the training data and testing data are clean, \textit{B}: we use the \ff{labels} from one of the six pathologists as training data (noisy) and the testing data is clean. %
The results are shown in Table~\ref{tab:gleason}. The comparison of A and B shows that inter-observer variability can cause a performance drop in medical image classification. All the methods show improvements over the baseline except F-correction. Our method shows the best \ff{results} among all the comparison methods.


\subsection{Ablation Analysis}
\label{sec:ab}

\para{Efficiency of key components.}
\xc{To demonstrate the effectiveness of our proposed method, we \ff{conducted} the ablation study using both the histopathologic cancer detection dataset and ISIC skin dataset. The results for each setting is shown in Table~\ref{tab:ab_hist}.
Our baseline is the ResNet34 trained by cross entropy loss and the result is provided in the first row of Table~\ref{tab:ab_hist}.}
We mainly \ff{analyzed} the efficiency of four components:  co-training with noisy label filter, self-ensemble, global inter-sample relationship alignment loss, and the local contrastive loss. 
From the ablation study, we \ff{observed} that each component plays its own role in a complementary way. 
Specifically, the co-training with a noisy label filter scheme improves the accuracy by around 1$\%$-4.7$\%$ compared to the baseline. It shows that the two networks cross updated by selected samples yield improvement over the baseline by reducing the domination of noisy labels and alleviating the sample selection bias. 
Then we \ff{evaluated} the efficiency of the self-ensemble setting. By using the self-ensemble model (mean teacher network), the accuracy improved $1.6 \% - 7.5 \%$ since the teacher network is a more robust moving average model of the student network. Hence, it is effective to adopt the self-ensemble model to select samples.
Lastly, we \ff{evaluated} the effectiveness of the two proposed self-supervised losses. As shown in line 4 and line 5 of Table \ref{tab:ab_hist}, only using the global inter-sample relationship alignment loss $l_{global}$ and the local contrastive loss $L_{local}$ achieved $1.7 \% -14.5 \%$ improvement on accuracy, indicates the effectiveness of these two losses. \ff{Our model utilized the two losses achieved the best performance} and the final accuracy is $2.6 \% - 15.6 \%$ higher than the baseline. It indicates that the representation learning capability of the network has been improved by training the network with global loss and local loss on the detected noisy labeled data.

\para{Different temperatures of contrastive loss.}
We \ff{analyzed} how the temperature hyperparameter in the local contrastive loss influences the network performance. We \ff{ranged} $t$ $\in$ [0.05, 0.1, 0.5, 1] for both datasets under the 10\% noise ratio setting. 
As shown in Table~\ref{tab:temperature}, the \ff{classification accuracy} generally improves the performance over \ff{the} baseline (83.01\% for skin and 88.95\% for lymph node) while not being very sensitive to the value of $t$.
\begin{figure}[t]
\begin{tabular}{cc}

  \hspace{-3mm}\includegraphics[width=0.5\columnwidth]{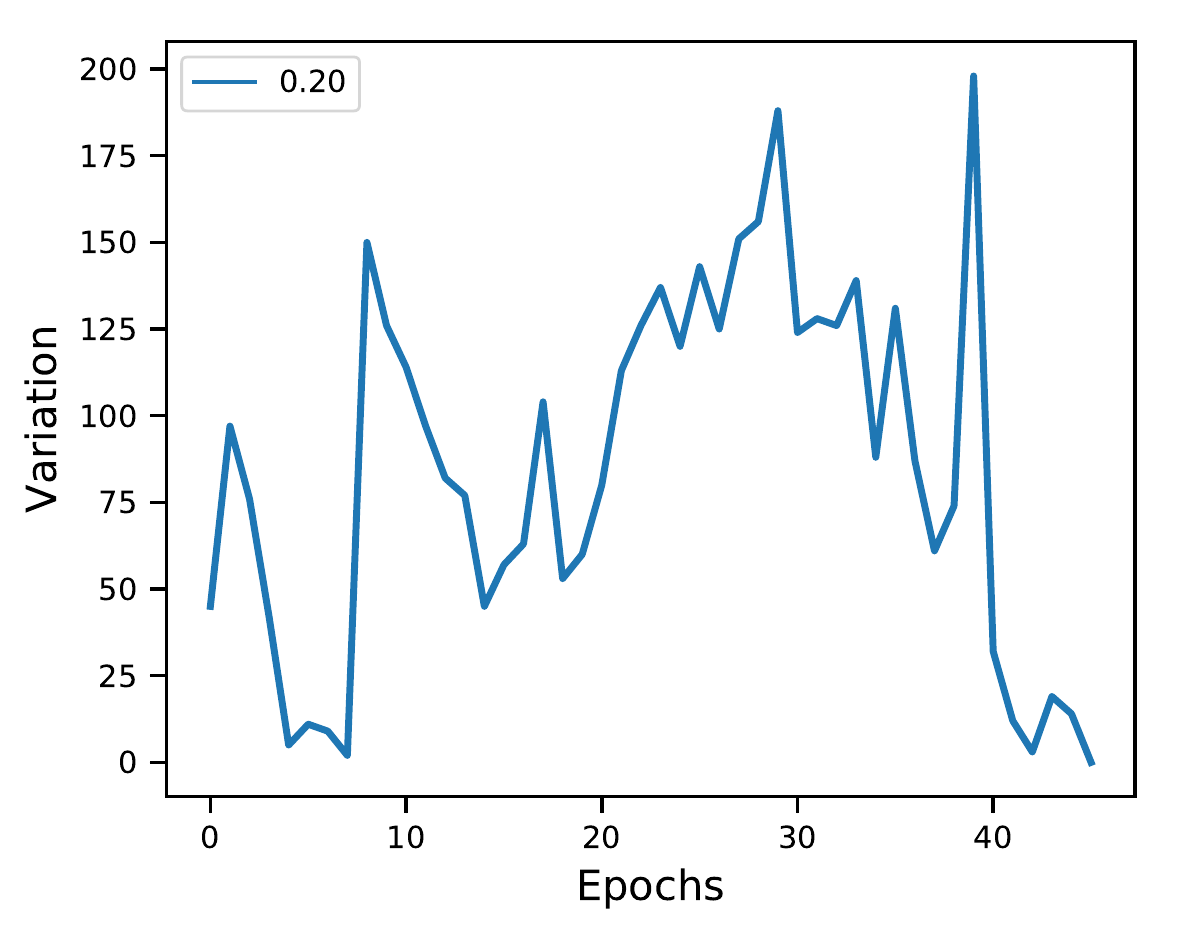} & \hspace{-7mm} \includegraphics[width=0.5\columnwidth]{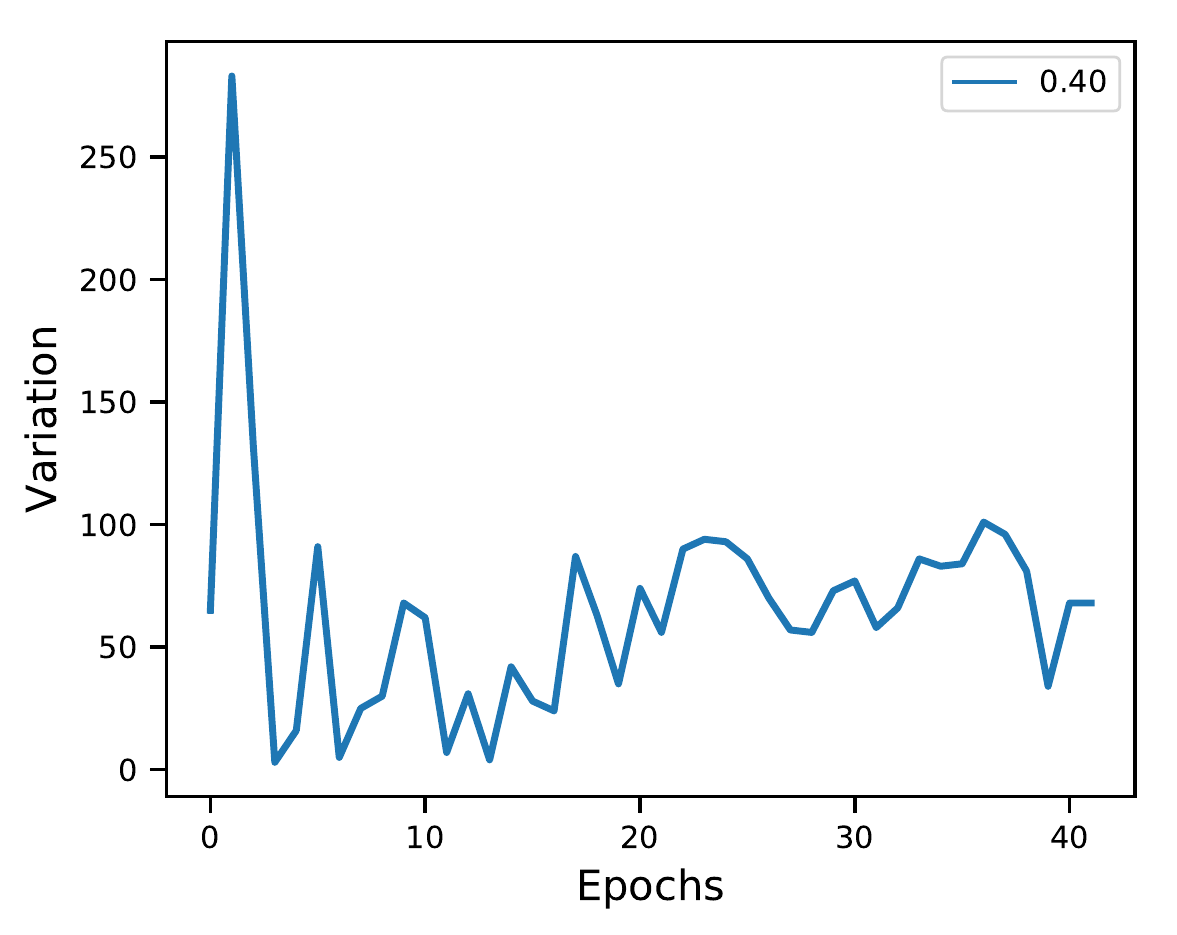}  \\
 
(a)  & (b)   \\
\end{tabular}
\caption{The variation of filtered labeled data number. 
(20 $\%$ and 40$\%$ noisy lymph node data)}
\label{fig:var}
\end{figure}

\para{Different weights of self-supervised loss.}
We also \ff{varied} the trade-off weight of the self-supervised loss and \ff{analyzed} its sensitivity to the hyperparameter of $\lambda$ in Eq.~\eqref{eq:total}. Specifically, we \ff{ranged} $\lambda \in [0,1,10,20]$ and \ff{observed} the classification \ff{results} on the histopathologic dataset. In Fig.~\ref{fig:auc}(a), the bar-plots \ff{presents} the mean of classification accuracy under all noise \xc{settings}. We \ff{observed} that our method generally improves the classification performance when $\lambda < 10$, \xcc{while a larger weight of the self-supervision will overwhelm the supervised one and lead to underfitting.}

\para{Noisy label detection.}
We first show the training loss distribution obtained by pure cross entropy loss and our methods at 10 epoch in Fig.~\ref{fig:hist_cm}. The training loss is easier to be divided into clean and noisy groups with our method, which is important for the noise label detection procedure. In Fig.~\ref{fig:auc}(b), we also plot the AUC value of the filtered clean samples at the last epoch. Our method can consistently select high-quality samples with a higher AUC value compared to using one network to select samples, demonstrating that the selection procedure in our method is more robust. The lymph node dataset was used in this part. \xc{The difference (variation) of filtered clean label sample numbers from two networks is presented in Fig. \ref{fig:var}. The two networks consistently selected a different number of samples, which shows the two networks have different learning abilities for noisy sample detection.}

\para{Learning curve of global alignment loss.}
In Fig.~\ref{fig:auc}(c), we plot the change of global inter-sample relationship with and without the global alignment loss. The results show that the loss of inter-sample relationships would not naturally converge during the network training when there is no explicit guidance. The global alignment loss can enhance the feature robustness by aligning the relationship.

%% file: 5-discussion.tex
\begin{figure}[t]
\begin{tabular}{ccc}
  \includegraphics[width=0.32\columnwidth]{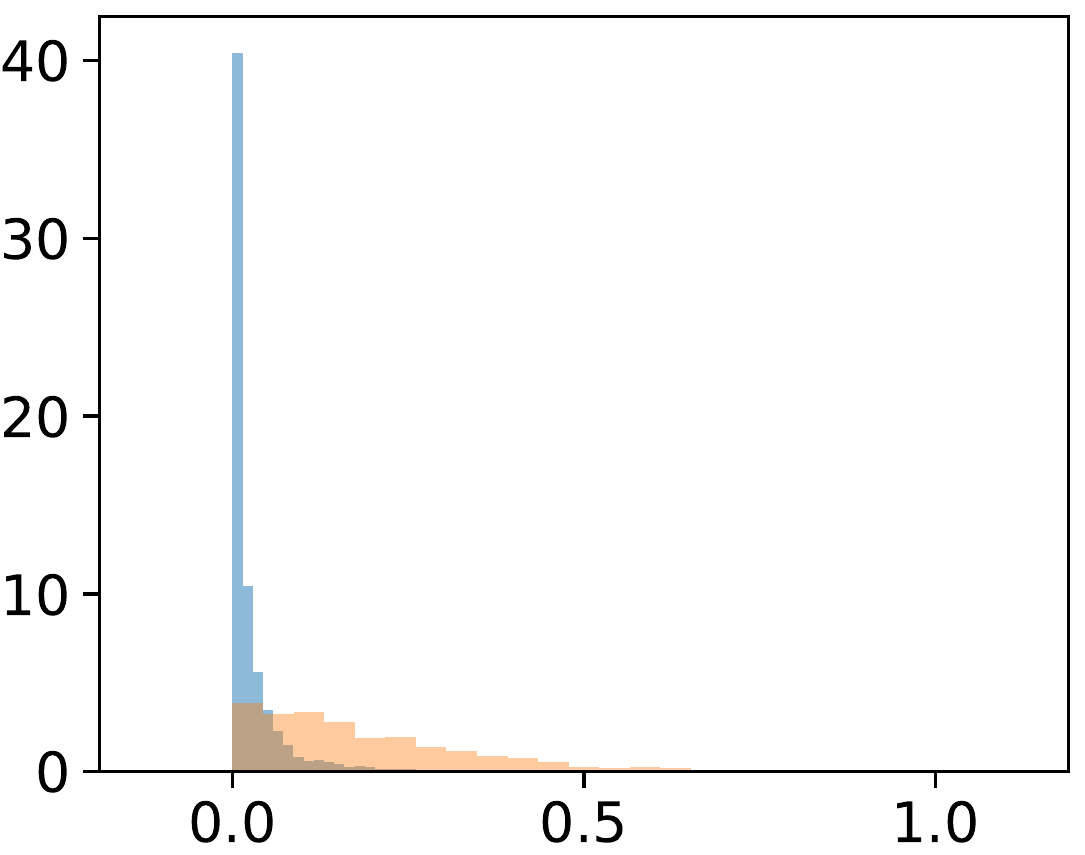} & \hspace{-5mm} \includegraphics[width=0.32\columnwidth]{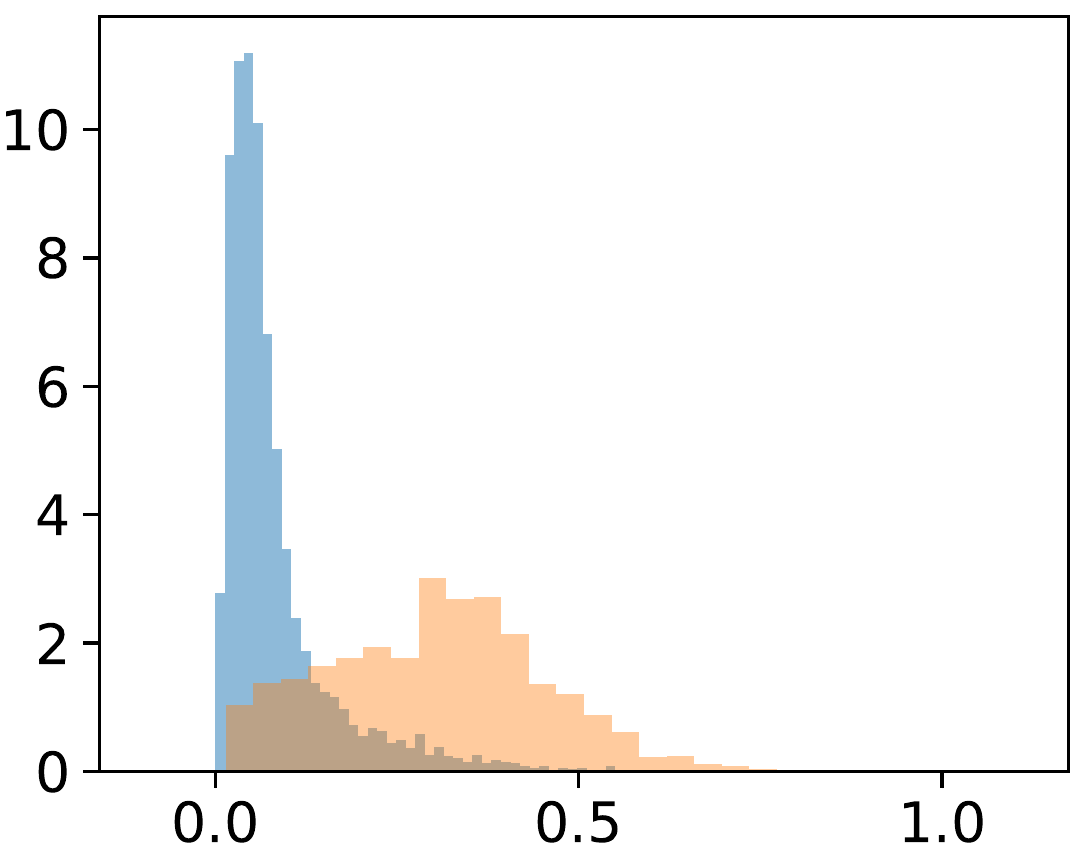} & \hspace{-3mm}\includegraphics[width=0.32\columnwidth]{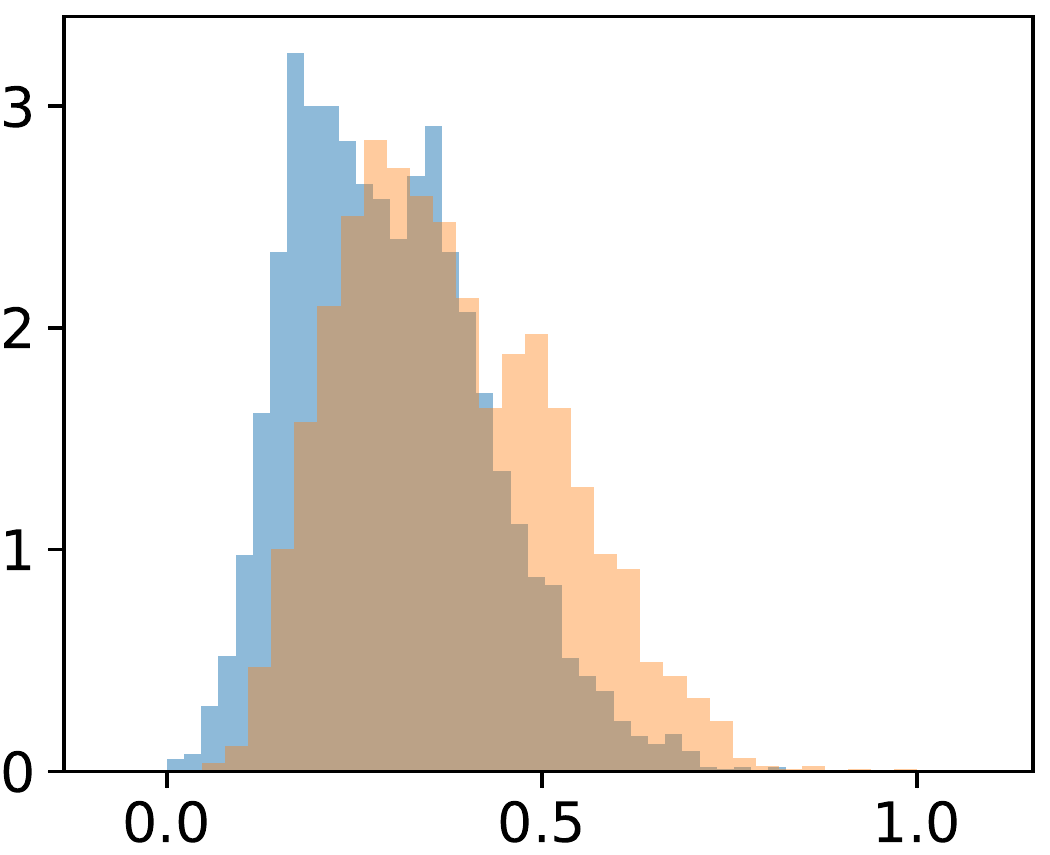} \\
 \includegraphics[width=0.32\columnwidth]{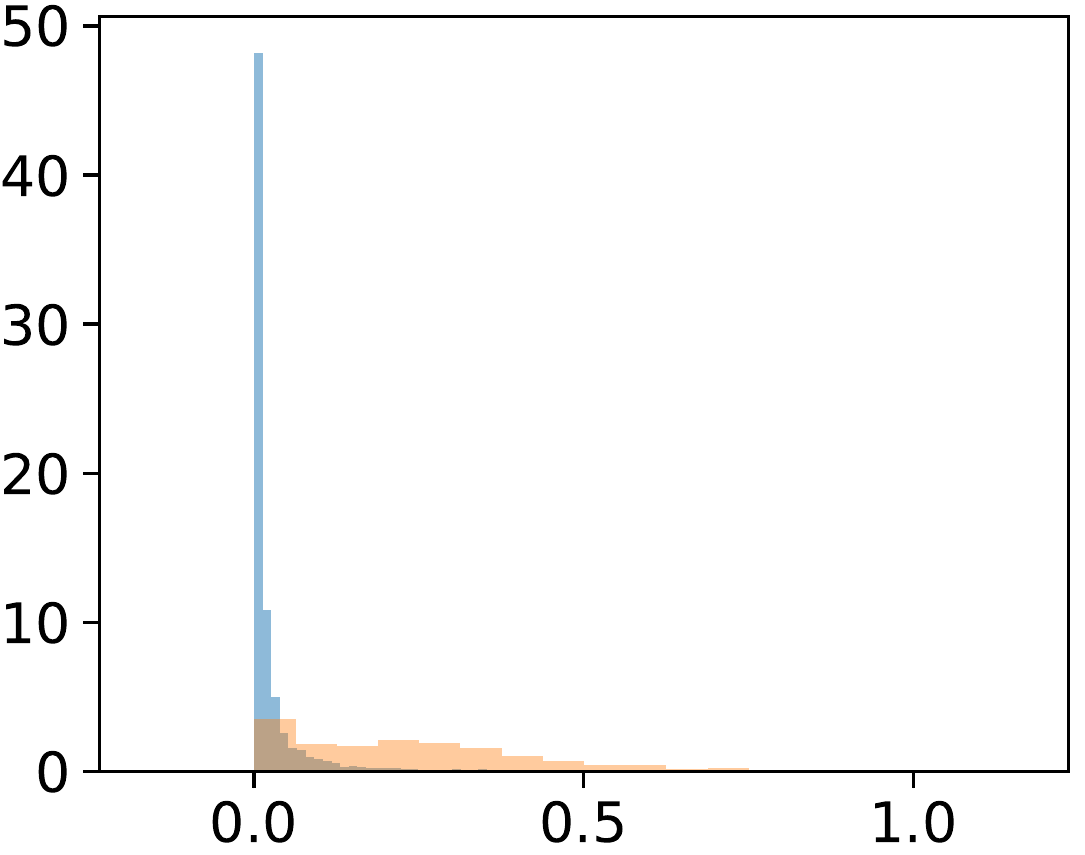} &\hspace{-5mm}   \includegraphics[width=0.32\columnwidth]{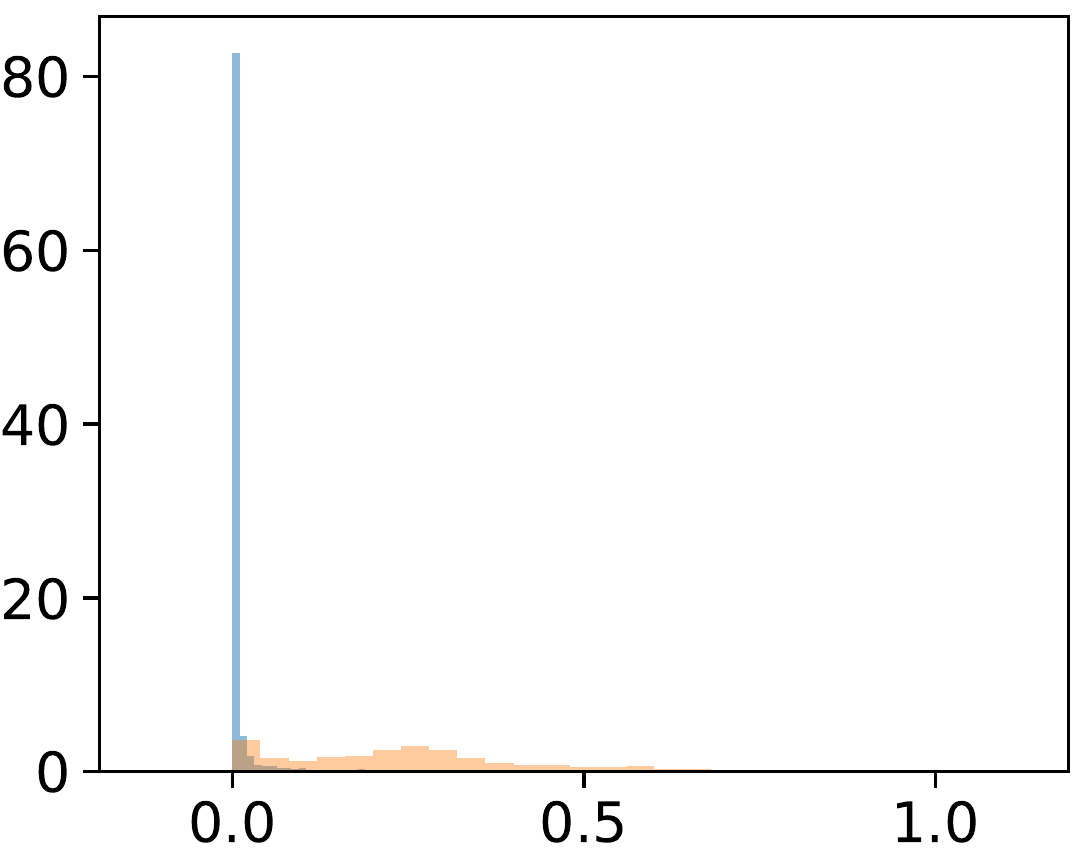} &
 \hspace{-3mm}\includegraphics[width=0.32\columnwidth]{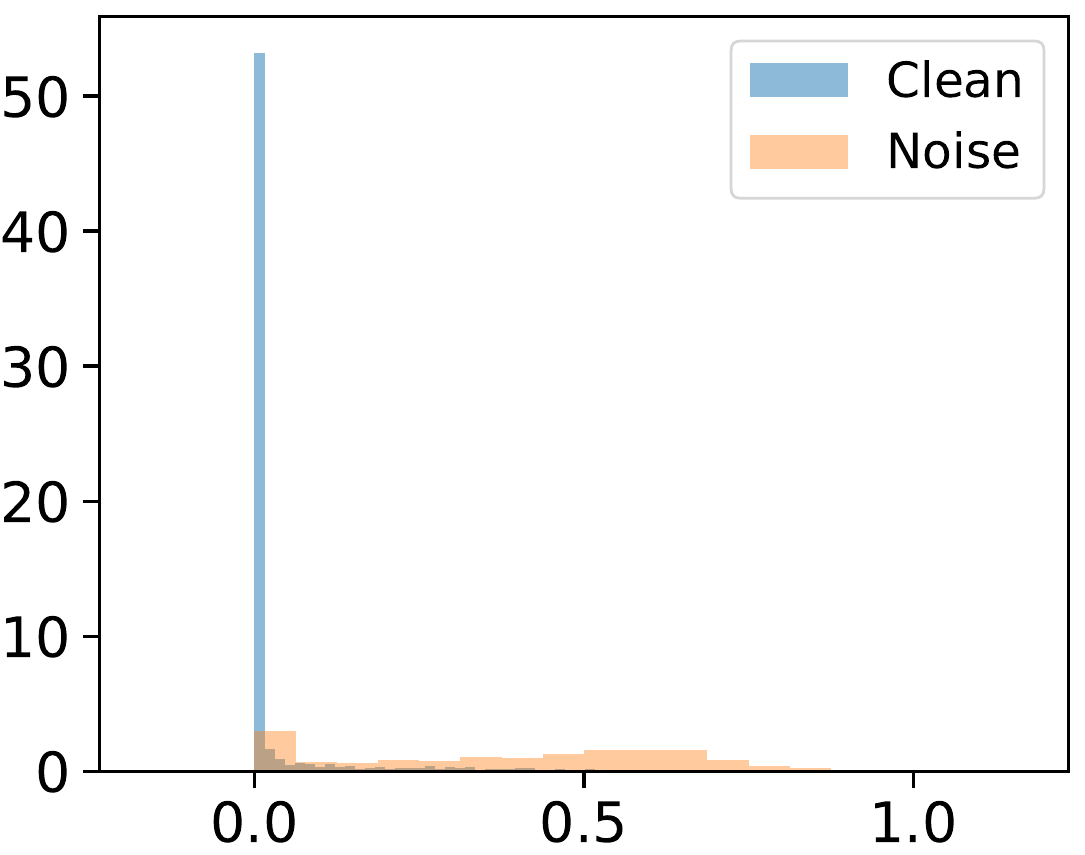}  \\
(a) $10\%$ & (b) $20\%$& (c) $40\%$ \\[2pt]
\end{tabular}
\caption{The comparison of the training loss distribution on histopathologic image patches. The first row represents the distribution trained with cross entropy while the second row represents the distribution of our method.}
\label{fig:hist_cm}
\end{figure}
\section{Discussion}
Deep neural networks usually require large-scale annotated training data, while the annotations of medical images usually are prone to experience and disagreement may appear between doctors. For diagnosis tasks in the medical image analysis field, the collection of large scale accurate annotated data is challenging due to the privacy policy and high work loading. 
Also, it is known that noisy labeled training data will hurt the performance of CNNs. 
Therefore, designing robust deep neural networks for medical image analysis is an appealing topic in the medical image analysis community. 
Many deep-learning-based methods have been developed for \xc{more robust training with corrupted training labels}~\cite{dgani2018training,xue2019robust,zhu2019pick}. 
However, these methods exist a strong accumulated error caused by sample selection bias which is coming from the wrongly selected samples. It will naturally influence the network performance and further decrease the quality of selected samples. 
On the other hand, the partial selection of training data would waste the valuable dataset during the training process. Furthermore, the hard sample is easy to be treated as the noisy label in medical image analysis, either discarding or re-labeling these data will lead to worse performance. 
In this work, through the self-ensemble co-training scheme with a robust noisy label filter and \ff{two tailored self-supervision loss} 
, a novel deep learning framework was proposed for the noisy labeled medical image dataset.

Inspired by the collaborative training scheme proposed by \cite{han2018co,malach2017decoupling}, a co-training network with a noisy label filter has been utilized to more explicitly enhance the network robustness and greatly alleviate the selection bias, as shown in Table~\ref{tab:ab_hist}. 
Moreover, we incorporate a novel self-supervised learning scheme towards the global relationship alignment of samples in each batch and local clustering of each sample, which could maximally utilize the detected noisy labeled dataset (\eg, noisy label or hard samples). The experimental results demonstrate that this strategy efficiently improves the utilization of training data and improves the model performance. 
Fig.~\ref{fig:hist_cm} shows that our self-ensemble network with two regularization losses could detect the noisy label more efficiently \ff{by properly dividing the noisy and clean dataset.}
\begin{table}[t]
\centering
\caption{Statistical test results for Gleason grading dataset.}
	\label{tab:p_gleason}
\begin{tabular}{c|cc}
\toprule[1.5pt]
              & A      & B      \\ \hline
Cross entropy & 9e-4 & 1e-4 \\
F-correction  & - & - \\
Mentornet     & 8e-4 & 1e-4 \\
Decoupling    & 2e-3 & 5e-4 \\
Co-teaching   & 9e-4 & 9e-4 \\
ELR           & 2e-3 &1e-4 \\ 
\bottomrule[1.5pt]
\end{tabular}
\end{table}
\begin{table}[t]
\centering
\caption{Statistical test results for NIH dataset.}
	\label{tab:p_nih}
\begin{tabular}{c|cc}
\toprule[1.5pt]
  & Pneumothorax & Nodule/Mass \\ \hline
p-value & 8e-4        & 8.3e-3 \\
\bottomrule[1.5pt]
\end{tabular}
\end{table}

\xcc{To evaluate the robustness of our method, it was tested on three types of noisy labels, \ie random noise, computer generated label, and inter-observer variability. For the random noise, we simulated two noisy datasets using the ISIC skin lesion data and the Kaggle lymph node data with different noisy ratios. The results suggested that our method can handle random noise even with very imbalanced data distribution. To better evaluate the proposed method in a real-world setting, we trained the model using the public NIH datasets with NLP mined label and tested it on a manually labeled Chest X-ray dataset to verify the effectiveness of our method in reality. Our method consistently shows better results than the baseline and other comparison methods, showing that the proposed method can handle the computer generated label with mild noise. The proposed method was also tested on the Gleason grading dataset with multi-annotators, which is labeled by six pathologists with different experiences. The dataset \ff{exists} great inter-observer disagreements, our experiment showed that when using STAPLE estimated label as clean testing label, different training labels showed significantly different test results. Our method achieved the best results among all the comparison methods with significant improvements ($p<0.05$, paired t-test), as shown in Table \ref{tab:p_gleason} and Table \ref{tab:p_nih}.}

Overall, we show that the co-training strategy and self-supervised regularization can work together to build a robust training scheme, which successfully selected the high quality data for training, as shown in Fig.~\ref{fig:auc}(b). Compared with other state-of-the-art methods, our method achieves significant improvement, no matter on a mild corrupted dataset or a heavily corrupted dataset.



\xcc{In this study, we didn't consider the disagreements between different annotators as a parameter during training, as recruiting annotations from multiple radiologists/pathologists is time-consuming in clinical practice. In the future, we would like to extend our method to sample dependent label noise, such as considering the grading of each doctor as a clue for the noisy labels. Secondly, even though the two-network setting could be effective for alleviating the selection bias, the computation cost doubles with two networks. One possible solution to alleviate the selection bias is to train the network in a meta-learning framework.}


\section{Conclusion}
In this paper, we present a global and local representation guided co-training strategy to address the challenging yet important noisy label issue for medical image analysis. The proposed method does not rely on refining or relabeling the noisy labeled data but employs two self-supervised losses to promote the learning of robust representation features. 
Our method can alleviate the wasting of data, including the hard samples, and avoid introducing new noise to training labels in a mild noisy dataset. The proposed framework can be easily extended to multi-class classification tasks and used in general classification networks for improving model robustness. 
\xcc{We extensively evaluated our method on four challenging medical image classification tasks with random noise, computer generated noise, and noise caused by inter-observer variability. Experimental results demonstrate the
effectiveness of our proposed framework in learning with the noisy label.}

%% file: 0-main.bbl
\begin{thebibliography}{10}
\providecommand{\url}[1]{#1}
\csname url@samestyle\endcsname
\providecommand{\newblock}{\relax}
\providecommand{\bibinfo}[2]{#2}
\providecommand{\BIBentrySTDinterwordspacing}{\spaceskip=0pt\relax}
\providecommand{\BIBentryALTinterwordstretchfactor}{4}
\providecommand{\BIBentryALTinterwordspacing}{\spaceskip=\fontdimen2\font plus
\BIBentryALTinterwordstretchfactor\fontdimen3\font minus
  \fontdimen4\font\relax}
\providecommand{\BIBforeignlanguage}[2]{{%
\expandafter\ifx\csname l@#1\endcsname\relax
\typeout{** WARNING: IEEEtran.bst: No hyphenation pattern has been}%
\typeout{** loaded for the language `#1'. Using the pattern for}%
\typeout{** the default language instead.}%
\else
\language=\csname l@#1\endcsname
\fi
#2}}
\providecommand{\BIBdecl}{\relax}
\BIBdecl

\bibitem{esteva2017dermatologist}
A.~Esteva, B.~Kuprel, R.~A. Novoa, J.~Ko, S.~M. Swetter, H.~M. Blau, and
  S.~Thrun, ``Dermatologist-level classification of skin cancer with deep
  neural networks,'' \emph{Nature}, vol. 542, no. 7639, p. 115, 2017.

\bibitem{setio2017validation}
A.~A.~A. Setio, A.~Traverso, T.~De~Bel, M.~S. Berens, C.~van~den Bogaard,
  P.~Cerello, H.~Chen, Q.~Dou, M.~E. Fantacci, B.~Geurts \emph{et~al.},
  ``Validation, comparison, and combination of algorithms for automatic
  detection of pulmonary nodules in computed tomography images: the luna16
  challenge,'' \emph{Medical image analysis}, vol.~42, pp. 1--13, 2017.

\bibitem{de2018clinically}
J.~De~Fauw, J.~R. Ledsam, B.~Romera-Paredes, S.~Nikolov, N.~Tomasev
  \emph{et~al.}, ``Clinically applicable deep learning for diagnosis and
  referral in retinal disease,'' \emph{Nature medicine}, vol.~24, no.~9, p.
  1342, 2018.

\bibitem{bejnordi2017diagnostic}
B.~E. Bejnordi, M.~Veta, P.~J. Van~Diest, B.~Van~Ginneken, N.~Karssemeijer,
  G.~Litjens, J.~A. Van Der~Laak, M.~Hermsen, Q.~F. Manson, M.~Balkenhol
  \emph{et~al.}, ``Diagnostic assessment of deep learning algorithms for
  detection of lymph node metastases in women with breast cancer,''
  \emph{Jama}, vol. 318, no.~22, pp. 2199--2210, 2017.

\bibitem{zhang2016understanding}
C.~Zhang, S.~Bengio, M.~Hardt, B.~Recht, and O.~Vinyals, ``{Understanding deep
  learning requires rethinking generalization},'' in \emph{ICLR}, 2017.

\bibitem{chen2019understanding}
P.~Chen, B.~Liao, G.~Chen, and S.~Zhang, ``Understanding and utilizing deep
  neural networks trained with noisy labels,'' \emph{arXiv preprint
  arXiv:1905.05040}, 2019.

\bibitem{dgani2018training}
Y.~Dgani, H.~Greenspan, and J.~Goldberger, ``Training a neural network based on
  unreliable human annotation of medical images,'' in \emph{Biomedical Imaging
  (ISBI 2018), 2018 IEEE 15th International Symposium on}.\hskip 1em plus 0.5em
  minus 0.4em\relax IEEE, 2018, pp. 39--42.

\bibitem{xue2019robust}
C.~Xue, Q.~Dou, X.~Shi, H.~Chen, and P.~A. Heng, ``Robust learning at noisy
  labeled medical images: Applied to skin lesion classification,'' in
  \emph{ISBI}, 2019.

\bibitem{zhu2019pick}
H.~Zhu, J.~Shi, and J.~Wu, ``Pick-and-learn: Automatic quality evaluation for
  noisy-labeled image segmentation,'' in \emph{International Conference on
  Medical Image Computing and Computer-Assisted Intervention}.\hskip 1em plus
  0.5em minus 0.4em\relax Springer, 2019, pp. 576--584.

\bibitem{shu2019lvc}
Y.~Shu, X.~Wu, and W.~Li, ``Lvc-net: Medical image segmentation with noisy
  label based on local visual cues,'' in \emph{International Conference on
  Medical Image Computing and Computer-Assisted Intervention}.\hskip 1em plus
  0.5em minus 0.4em\relax Springer, 2019, pp. 558--566.

\bibitem{xue2020cascaded}
C.~Xue, Q.~Deng, X.~Li, Q.~Dou, and P.-A. Heng, ``Cascaded robust learning at
  imperfect labels for chest x-ray segmentation,'' in \emph{International
  Conference on Medical Image Computing and Computer-Assisted
  Intervention}.\hskip 1em plus 0.5em minus 0.4em\relax Springer, 2020, pp.
  579--588.

\bibitem{goldberger2016training}
J.~Goldberger and E.~Ben-Reuven, ``{Training deep neural-networks using a noise
  adaptation layer},'' in \emph{ICLR}, 2017.

\bibitem{patrini2017making}
G.~Patrini, A.~Rozza, A.~K. Menon, R.~Nock, and L.~Qu, ``{Making deep neural
  networks robust to label noise: A loss correction approach},'' in
  \emph{CVPR}, 2017, pp. 2233--2241.

\bibitem{zhengerror}
S.~Zheng, P.~Wu, A.~Goswami, M.~Goswami, D.~Metaxas, and C.~Chen,
  ``Error-bounded correction of noisy labels,'' in \emph{International
  Conference on Machine Learning}.\hskip 1em plus 0.5em minus 0.4em\relax PMLR,
  2020, pp. 11\,447--11\,457.

\bibitem{jiang2017mentornet}
L.~Jiang, Z.~Zhou, T.~Leung, L.-J. Li, and L.~Fei-Fei, ``{MentorNet:
  Regularizing very deep neural networks on corrupted labels},'' in
  \emph{ICML}, 2018.

\bibitem{han2018co}
B.~Han, Q.~Yao, X.~Yu, G.~Niu, M.~Xu, W.~Hu, I.~Tsang, and M.~Sugiyama,
  ``Co-teaching: Robust training of deep neural networks with extremely noisy
  labels,'' in \emph{Advances in neural information processing systems}, 2018,
  pp. 8527--8537.

\bibitem{hendrycks2018using}
D.~Hendrycks, M.~Mazeika, D.~Wilson, and K.~Gimpel, ``Using trusted data to
  train deep networks on labels corrupted by severe noise,'' in \emph{Advances
  in neural information processing systems}, 2018, pp. 10\,456--10\,465.

\bibitem{jiang2018hyperspectral}
J.~Jiang, J.~Ma, Z.~Wang, C.~Chen, and X.~Liu, ``Hyperspectral image
  classification in the presence of noisy labels,'' \emph{IEEE Transactions on
  Geoscience and Remote Sensing}, vol.~57, no.~2, pp. 851--865, 2018.

\bibitem{xia2019anchor}
X.~Xia, T.~Liu, N.~Wang, B.~Han, C.~Gong, G.~Niu, and M.~Sugiyama, ``Are anchor
  points really indispensable in label-noise learning?'' in \emph{Advances in
  Neural Information Processing Systems}, 2019, pp. 6838--6849.

\bibitem{yao2020dual}
Y.~Yao, T.~Liu, B.~Han, M.~Gong, J.~Deng, G.~Niu, and M.~Sugiyama, ``Dual t:
  Reducing estimation error for transition matrix in label-noise learning,''
  \emph{Advances in Neural Information Processing Systems}, vol.~33, 2020.

\bibitem{sukhbaatar2014training}
S.~Sukhbaatar, J.~Bruna, M.~Paluri, L.~Bourdev, and R.~Fergus, ``Training
  convolutional networks with noisy labels,'' \emph{arXiv preprint
  arXiv:1406.2080}, 2014.

\bibitem{tanaka2018joint}
D.~Tanaka, D.~Ikami, T.~Yamasaki, and K.~Aizawa, ``{Joint optimization
  framework for learning with noisy labels},'' in \emph{CVPR}, 2018.

\bibitem{ren2018learning}
M.~Ren, W.~Zeng, B.~Yang, and R.~Urtasun, ``{Learning to reweight examples for
  robust deep learning},'' in \emph{ICML}, 2018.

\bibitem{nguyen2019self}
D.~T. Nguyen, C.~K. Mummadi, T.~P.~N. Ngo, T.~H.~P. Nguyen, L.~Beggel, and
  T.~Brox, ``Self: Learning to filter noisy labels with self-ensembling,''
  \emph{arXiv preprint arXiv:1910.01842}, 2019.

\bibitem{li2020dividemix}
J.~Li, R.~Socher, and S.~C. Hoi, ``Dividemix: Learning with noisy labels as
  semi-supervised learning,'' \emph{arXiv preprint arXiv:2002.07394}, 2020.

\bibitem{ding2018semi}
Y.~Ding, L.~Wang, D.~Fan, and B.~Gong, ``A semi-supervised two-stage approach
  to learning from noisy labels,'' in \emph{2018 IEEE Winter Conference on
  Applications of Computer Vision (WACV)}.\hskip 1em plus 0.5em minus
  0.4em\relax IEEE, 2018, pp. 1215--1224.

\bibitem{nguyen2019robust}
D.~T. Nguyen, T.-P.-N. Ngo, Z.~Lou, M.~Klar, L.~Beggel, and T.~Brox, ``Robust
  learning under label noise with iterative noise-filtering,'' \emph{arXiv
  preprint arXiv:1906.00216}, 2019.

\bibitem{liu2020earlylearning}
S.~Liu, J.~Niles-Weed, N.~Razavian, and C.~Fernandez-Granda, ``Early-learning
  regularization prevents memorization of noisy labels,'' \emph{Advances in
  Neural Information Processing Systems}, vol.~33, 2020.

\bibitem{min2019two}
S.~Min, X.~Chen, Z.-J. Zha, F.~Wu, and Y.~Zhang, ``A two-stream mutual
  attention network for semi-supervised biomedical segmentation with noisy
  labels,'' in \emph{Proceedings of the AAAI Conference on Artificial
  Intelligence}, vol.~33, no.~01, 2019, pp. 4578--4585.

\bibitem{le2019pancreatic}
H.~Le, D.~Samaras, T.~Kurc, R.~Gupta, K.~Shroyer, and J.~Saltz, ``Pancreatic
  cancer detection in whole slide images using noisy label annotations,'' in
  \emph{International Conference on Medical Image Computing and
  Computer-Assisted Intervention}.\hskip 1em plus 0.5em minus 0.4em\relax
  Springer, 2019, pp. 541--549.

\bibitem{wang2018iterative}
Y.~Wang, W.~Liu, X.~Ma, J.~Bailey, H.~Zha, L.~Song, and S.-T. Xia, ``Iterative
  learning with open-set noisy labels,'' in \emph{Proceedings of the IEEE
  Conference on Computer Vision and Pattern Recognition}, 2018, pp. 8688--8696.

\bibitem{codella2018skin}
N.~C. Codella~et al., ``Skin lesion analysis toward melanoma detection: A
  challenge at the 2017 international symposium on biomedical imaging (isbi),
  hosted by the international skin imaging collaboration (isic),'' in
  \emph{Biomedical Imaging (ISBI 2018), 2018 IEEE 15th International Symposium
  on}.\hskip 1em plus 0.5em minus 0.4em\relax IEEE, 2018, pp. 168--172.

\bibitem{demyanov2016classification}
S.~Demyanov, R.~Chakravorty, M.~Abedini, A.~Halpern, and R.~Garnavi,
  ``Classification of dermoscopy patterns using deep convolutional neural
  networks,'' in \emph{2016 IEEE 13th International Symposium on Biomedical
  Imaging (ISBI)}.\hskip 1em plus 0.5em minus 0.4em\relax IEEE, 2016, pp.
  364--368.

\bibitem{kawahara2016deep}
J.~Kawahara, A.~BenTaieb, and G.~Hamarneh, ``Deep features to classify skin
  lesions,'' in \emph{2016 IEEE 13th International Symposium on Biomedical
  Imaging (ISBI)}.\hskip 1em plus 0.5em minus 0.4em\relax IEEE, 2016, pp.
  1397--1400.

\bibitem{premaladha2016novel}
J.~Premaladha and K.~Ravichandran, ``Novel approaches for diagnosing melanoma
  skin lesions through supervised and deep learning algorithms,'' \emph{Journal
  of medical systems}, vol.~40, no.~4, p.~96, 2016.

\bibitem{yu2016automated}
L.~Yu, H.~Chen, Q.~Dou, J.~Qin, and P.-A. Heng, ``Automated melanoma
  recognition in dermoscopy images via very deep residual networks,''
  \emph{IEEE transactions on medical imaging}, vol.~36, no.~4, pp. 994--1004,
  2016.

\bibitem{wang2016deep}
D.~Wang, A.~Khosla, R.~Gargeya, H.~Irshad, and A.~H. Beck, ``Deep learning for
  identifying metastatic breast cancer,'' \emph{arXiv preprint
  arXiv:1606.05718}, 2016.

\bibitem{kong2017cancer}
B.~Kong, X.~Wang, Z.~Li, Q.~Song, and S.~Zhang, ``Cancer metastasis detection
  via spatially structured deep network,'' in \emph{International Conference on
  Information Processing in Medical Imaging}.\hskip 1em plus 0.5em minus
  0.4em\relax Springer, 2017, pp. 236--248.

\bibitem{courtiol2018classification}
P.~Courtiol, E.~W. Tramel, M.~Sanselme, and G.~Wainrib, ``Classification and
  disease localization in histopathology using only global labels: A
  weakly-supervised approach,'' \emph{arXiv preprint arXiv:1802.02212}, 2018.

\bibitem{liu2017detecting}
Y.~Liu, K.~Gadepalli, M.~Norouzi, G.~E. Dahl, T.~Kohlberger, A.~Boyko,
  S.~Venugopalan, A.~Timofeev, P.~Q. Nelson, G.~S. Corrado \emph{et~al.},
  ``Detecting cancer metastases on gigapixel pathology images,'' \emph{arXiv
  preprint arXiv:1703.02442}, 2017.

\bibitem{cruz2017accurate}
A.~Cruz-Roa, H.~Gilmore, A.~Basavanhally, M.~Feldman, S.~Ganesan, N.~N. Shih,
  J.~Tomaszewski, F.~A. Gonz{\'a}lez, and A.~Madabhushi, ``Accurate and
  reproducible invasive breast cancer detection in whole-slide images: A deep
  learning approach for quantifying tumor extent,'' \emph{Scientific reports},
  vol.~7, p. 46450, 2017.

\bibitem{malach2017decoupling}
E.~Malach and S.~Shalev-Shwartz, ``Decoupling" when to update" from" how to
  update",'' in \emph{Advances in Neural Information Processing Systems}, 2017,
  pp. 960--970.

\bibitem{chen2020simple}
T.~Chen, S.~Kornblith, M.~Norouzi, and G.~Hinton, ``A simple framework for
  contrastive learning of visual representations,'' \emph{arXiv preprint
  arXiv:2002.05709}, 2020.

\bibitem{avni2010x}
U.~Avni, H.~Greenspan, E.~Konen, M.~Sharon, and J.~Goldberger, ``X-ray
  categorization and retrieval on the organ and pathology level, using
  patch-based visual words,'' \emph{IEEE Transactions on Medical Imaging},
  vol.~30, no.~3, pp. 733--746, 2010.

\bibitem{tarvainen2017mean}
A.~Tarvainen and H.~Valpola, ``Mean teachers are better role models:
  Weight-averaged consistency targets improve semi-supervised deep learning
  results,'' in \emph{Advances in neural information processing systems}, 2017,
  pp. 1195--1204.

\bibitem{berthelot2019mixmatch}
D.~Berthelot, N.~Carlini, I.~Goodfellow, N.~Papernot, A.~Oliver, and C.~A.
  Raffel, ``Mixmatch: A holistic approach to semi-supervised learning,'' in
  \emph{Advances in Neural Information Processing Systems}, 2019, pp.
  5049--5059.

\bibitem{veeling2018rotation}
B.~S. Veeling, J.~Linmans, J.~Winkens, T.~Cohen, and M.~Welling, ``Rotation
  equivariant cnns for digital pathology,'' in \emph{International Conference
  on Medical image computing and computer-assisted intervention}.\hskip 1em
  plus 0.5em minus 0.4em\relax Springer, 2018, pp. 210--218.

\bibitem{wang2017chestx}
X.~Wang, Y.~Peng, L.~Lu, Z.~Lu, M.~Bagheri, and R.~M. Summers, ``Chestx-ray8:
  Hospital-scale chest x-ray database and benchmarks on weakly-supervised
  classification and localization of common thorax diseases,'' in
  \emph{Proceedings of the IEEE conference on computer vision and pattern
  recognition}, 2017, pp. 2097--2106.

\bibitem{majkowska2020chest}
A.~Majkowska, S.~Mittal, D.~F. Steiner, J.~J. Reicher, S.~M. McKinney, G.~E.
  Duggan, K.~Eswaran, P.-H. Cameron~Chen, Y.~Liu, S.~R. Kalidindi
  \emph{et~al.}, ``Chest radiograph interpretation with deep learning models:
  assessment with radiologist-adjudicated reference standards and
  population-adjusted evaluation,'' \emph{Radiology}, vol. 294, no.~2, pp.
  421--431, 2020.

\bibitem{nir2018automatic}
G.~Nir, S.~Hor, D.~Karimi, L.~Fazli, B.~F. Skinnider, P.~Tavassoli, D.~Turbin,
  C.~F. Villamil, G.~Wang, R.~S. Wilson \emph{et~al.}, ``Automatic grading of
  prostate cancer in digitized histopathology images: Learning from multiple
  experts,'' \emph{Medical image analysis}, vol.~50, pp. 167--180, 2018.

\bibitem{karimi2019deep}
D.~Karimi, G.~Nir, L.~Fazli, P.~C. Black, L.~Goldenberg, and S.~E. Salcudean,
  ``Deep learning-based gleason grading of prostate cancer from histopathology
  images—role of multiscale decision aggregation and data augmentation,''
  \emph{IEEE journal of biomedical and health informatics}, vol.~24, no.~5, pp.
  1413--1426, 2019.

\bibitem{karimi2020deep}
D.~Karimi, H.~Dou, S.~K. Warfield, and A.~Gholipour, ``Deep learning with noisy
  labels: Exploring techniques and remedies in medical image analysis,''
  \emph{Medical Image Analysis}, vol.~65, p. 101759, 2020.

\bibitem{warfield2004simultaneous}
S.~K. Warfield, K.~H. Zou, and W.~M. Wells, ``Simultaneous truth and
  performance level estimation (staple): an algorithm for the validation of
  image segmentation,'' \emph{IEEE transactions on medical imaging}, vol.~23,
  no.~7, pp. 903--921, 2004.

\end{thebibliography}
